\documentclass[prd, superscriptaddress, preprintnumbers,showpacs]{revtex4}
\usepackage{ifpdf}
\usepackage{graphicx}
\usepackage{amsmath}
\usepackage{multirow}
\usepackage{bm}
\usepackage{color}
\usepackage{ulem}
\newcommand{\sign}{\mbox{sgn}}

\begin{document}
\normalem
\title{Coulomb interaction and magnetic catalysis in the quantum Hall effect in graphene}
\date{\today}

\preprint{UWO-TH-11/6}

\author{E.V. Gorbar}
\email{gorbar@bitp.kiev.ua}
\affiliation{Bogolyubov Institute for Theoretical Physics, 03680, Kiev, Ukraine}

\author{V.P. Gusynin}
\email{vgusynin@bitp.kiev.ua}
\affiliation{Bogolyubov Institute for Theoretical Physics, 03680, Kiev, Ukraine}

\author{V.A. Miransky}
\email{vmiransk@uwo.ca}
\affiliation{Department of Applied Mathematics, University of Western Ontario,
London, Ontario N6A 5B7, Canada}

\author{I.A. Shovkovy}
\email{igor.shovkovy@asu.edu}
\affiliation{Department of Applied Sciences and Mathematics, Arizona State
University, Mesa, Arizona 85212, USA}

\begin{abstract}
The dynamics of symmetry breaking responsible for lifting the degeneracy of the 
Landau levels (LLs) in the integer quantum Hall (QH) effect in graphene is studied in a low-energy 
model with the Coulomb interaction. The gap equation for Dirac quasiparticles is analyzed 
for both the lowest and higher LLs, taking into account the LLs mixing.
It is shown that the characteristic feature of the long-range Coulomb
interaction is the dependence of the gap parameters on the LL index
$n$ (``running'' gaps). The renormalization (running) of the Fermi velocity as a function
of $n$ is also studied. The solutions of the gap equation reproduce correctly the experimentally
observed integer QH plateaus in graphene in strong magnetic fields.
\end{abstract}

\pacs{73.22.Pr, 71.70.Di, 71.70.-d}

\maketitle

\section{Introduction}
\label{1}

As it is well known, the low-energy dynamics of electrons in graphene \cite{Geim2004Science}
is described by the Dirac
equation in (2+1) dimensions \cite{Semenoff1984PRL}. Perhaps the most direct confirmation of
the pseudorelativistic character of electron motion in graphene is given by the experimental
observation \cite{Geim2005Nature,Kim2005Nature} of the anomalous quantum Hall (QH) effect
theoretically predicted in Refs.~\cite{Ando2002,Gusynin2005PRL,Peres2005}. The anomalous
QH plateaus are observed at the filling factors $\nu = \pm 4(n + 1/2)$, where $n=0,1,2,\ldots$
is the Landau level (LL) index. The factor 4 in the filling factor is due to a fourfold (spin and
sublattice-valley) degeneracy of each QH state in graphene. The presence of the anomalous
(from the viewpoint of more standard condensed matter systems) term $1/2$ in the filling factor
unmistakably reveals the relativistic-like character of electron motion
in graphene \cite{Semenoff1984PRL,Haldane1988PRL,Khveshchenko2001PRL,Gorbar2002PRB,Gusynin2004PRB}.

The later experiments \cite{Zhang2006,Jiang2007} in strong magnetic fields ($B \gtrsim 20\, \mbox{T}$)
observed new QH plateaus, with integer filling factors  $\nu = 0, \pm 1$, and $\pm 4$. The more recent
experiments \cite{Andrei2009,Bolotin2009} discovered additional plateaus, with $\nu = \pm 3$ and $\nu \pm 1/3$.
While the latter corresponds to the fractional QH effect, the plateaus with $\nu = 0, \pm 1, \pm 4$, and
$\nu = \pm 3$ are intimately connected with a breakdown of the $U(4)$ symmetry of the low-energy
effective quasiparticle Hamiltonian in graphene (connected with the spin and sublattice-valley degeneracy
mentioned above) \cite{Gorbar2002PRB}.
Strictly speaking, because of the Zeeman effect, the $U(4)$ is reduced to
a $U_{\uparrow}(2) \times U_{\downarrow}(2)$, with $U(2)_s$ being the sublattice-valley symmetry at a
fixed spin ($s=\uparrow$ or $s=\downarrow$). However, taking into account that the
Zeeman interaction is rather weak for realistic magnetic fields, the $U(4)$ is a good approximate
symmetry guaranteeing that at weak magnetic fields only the QH plateaus with the filling factors
$\nu=\pm 4(n+1/2)$ appear. The observed new QH plateaus $\nu = 0, \pm 1, \pm 4$, and
$\nu = \pm 3$ occur clearly due to the electron-electron interaction leading to (quasi-)spontaneous
$U(4)$ symmetry breaking that removes the degeneracy of the $n=0$ and $n=1$ LLs.

For the description of the new QH plateaus, the following two theoretical scenarios were
suggested. One of them is the QH ferromagnetism
(QHF) \cite{Nomura2006PRL,Goerbig2006,Alicea2006PRB,Sheng2007}, whose order parameters
are the spin and valley charge densities (the dynamics of a Zeeman spin splitting enhancement
considered in Ref.~\cite{Abanin2006PRL} is intimately connected with the QHF). This scenario
is related to the theory of exchange-driven spin splitting of LLs \cite{Fogler1995}. The second
one is the magnetic catalysis (MC) scenario, whose order parameters are excitonic condensates, responsible
for the generation of the Dirac masses of charge carriers \cite{Gusynin2006catalysis,Herbut2006,Fuchs2006,Ezawa2006}.
The essence of the magnetic catalysis, which is connected with the effective dimensional reduction
in the dynamics of charged fermions in an external magnetic field, was revealed in
Ref.~\cite{Gusynin1994PRL}. It was first applied for a single layer of graphite in
Refs.~\cite{Khveshchenko2001PRL,Gorbar2002PRB}.

One may think that the QHF and MC order parameters should compete with each other. However, the
analysis in an effective model with the local four-fermion interaction performed in Ref.~\cite{Gorbar2008PRB}
showed that these two sets of the order parameters necessarily coexist (this feature has been 
recently discussed also in Ref.~\cite{Semenoff2011}). This fact strongly indicates that 
these two sets of the order parameters have a common dynamical origin, i.e., they are two sides of the 
same coin. Their simultaneous consideration qualitatively reproduces all the QH plateaus observed 
experimentally in strong magnetic fields.

Certainly, it would be important to extend the analysis of Ref.~\cite{Gorbar2008PRB}
to the case of realistic long-range Coulomb interactions. In the present paper, the
dynamics of $U(4)$ symmetry breaking responsible for the appearance of the QH plateaus
with $\nu = 0, \pm 1$ in the lowest LL and with $\pm 3, \pm 4$, and $\pm 5$ in the $n = \pm 1$ 
LL is studied in a low-energy model with the Coulomb interaction. While the symmetric structure
of the solutions descibing these plateaus is similar to that of the solutions in the model
with the local four-fermion interaction \cite{Gorbar2008PRB}, there are essential
qualitative differences between them. In particular, the long-range Coulomb
interaction leads to the decrease of the gap parameters with increasing the LL index
$n$ (``running'' gaps).

Recently, the dynamics with the Coulomb interaction in the $\nu =0$ quantum Hall state 
was studied in Refs.~\cite{Kharitonov2011,Semenoff2011} by utilizing different 
approaches than the present one. In these papers, the important role of the Landau 
levels mixing effects was also revealed. Note that these mixing effects are important 
both in models with long-range interactions and those with short-range ones, as the 
model considered in Ref.~\cite{Gorbar2008PRB}.

The rest of the paper is organized as follows. In Sec.~\ref{secII}, the general features of
the model, in particular the structure of the order parameters, are described. The Dirac
quasiparticle propagator and energy dispersion relations are considered in Sec.~\ref{secIII}.
The Schwinger--Dyson (gap) equation for the quasiparticle propagator is derived
in Sec.~\ref{secIV}. In Sec.~\ref{secV}, we present our numerical results, which include
(i) the renormalization of the LL-dependent Fermi velocity parameter, (ii) the solutions
of the gap equation at the lowest Landau level (LLL) and (iii) the solutions of the gap equation at
the $n=1$ LL, respectively. The general discussion of the main results and a summary
are given in Sec.~\ref{Discussion}. Several appendices at the end of the paper contain
some of our derivations and technical details used in the main text.

Note that in most parts of the paper, we use the units with $\hbar=1$ and $k_B=1$.

\section{Model}
\label{secII}

The low-energy quasiparticle excitations in graphene are conveniently described in
terms of a four-component Dirac spinor $\Psi_s = \left( \psi_{KAs},\psi_{KBs},\psi_{K^\prime Bs},
\psi_{K^\prime As}\right)$ which combines the Bloch states with spin indices
$s=\uparrow$ or $s=\downarrow$ on the two different sublattices ($A,B$) of the hexagonal graphene
lattice and with momenta near the two inequivalent points ($K,K^\prime$) at the
opposite corners of the two-dimensional Brillouin zone. The free quasiparticle
Hamiltonian has a pseudorelativistic form with the Fermi
velocity $v_F\approx 10^6 \mbox{m/s}$ playing the role of the speed of light
\begin{equation}
 H_0=v_F\int d^2\mathbf{r}\,\overline{\Psi}_s\left(\gamma^1\pi_x+
\gamma^2\pi_y\right)\Psi_s, \label{free-hamiltonian}
\end{equation}
where $\mathbf{r} =(x,y)$, $\overline{\Psi}_{s}=\Psi^\dagger_{s}\gamma^0$ is the Dirac conjugated
spinor and summation over spin is understood. In Eq.~(\ref{free-hamiltonian}), $\gamma^\mu$
with $\mu=0,1,2$ are $4\times4$ gamma matrices belonging to a reducible representation
$\gamma^\mu=\tilde{\tau}_3\otimes (\tau_3,i\tau_2,-i\tau_1)$ where the Pauli matrices
$\tilde{\tau},\tau$ act in the subspaces of the valley ($K,K^\prime$) and sublattice ($A,B$)
indices, respectively. (For the Dirac $\gamma$-matrices, we use the same representation as
in Ref.~\cite{Gusynin2006catalysis}, Appendix~C.) The $\gamma$-matrices satisfy the usual
anticommutation relations $\left\{\gamma^\mu,\gamma^\nu\right\}=2g^{\mu\nu}$,
$g^{\mu\nu}= {\rm diag}\,(1,-1,-1)\,, \mu,\nu=0,1,2$. The canonical momentum
$\bm{\pi} \equiv (\pi_x, \pi_y)= -i\hbar\bm{\nabla}+ e\bm{A}/c$ includes the vector
potential $\bm{A}$ corresponding to a magnetic field $\bm{B}_{\perp}$, which is the
component of the external magnetic field $\bm{B}$ orthogonal to the graphene
$xy$-plane.

The Coulomb interaction term has the form
\begin{eqnarray} H_{C} &=& \frac{1}{2}\int d^2{\bf
r}d^2\mathbf{r}^\prime {\Psi}^{\dagger}_{s}(\mathbf{r})
\Psi_s(\mathbf{r})U_{C}(\mathbf{r}-\mathbf{r}^\prime)   % \nonumber\\ &\times&
{\Psi}^{\dagger}_{s^\prime}(\mathbf{r}^\prime) \Psi_{s^\prime}(\mathbf{r}^\prime),
\label{Coulomb}
\end{eqnarray}
where $U_{C}(\mathbf{r})$ is the Coulomb potential in a magnetic field. The corresponding
potential with the polarization effects taken into account was represented, for example,
in Eq.~(46) of Ref.~\cite{Gorbar2002PRB}. The Hamiltonian
$H = H_0+H_{C}$ possesses the $U(4)$ symmetry discussed above.
The electron chemical potential $\mu$ is introduced through adding the term
$-\mu \Psi^{\dagger}\Psi$ in $H$ (this term preserves the $U(4)$ symmetry).
The Zeeman term $\mu_BB\Psi^{\dagger} \sigma_3 \Psi$, where $B \equiv |\bm{B}|$ and
$\mu_B=e\hbar/(2mc)$ is the Bohr magneton, and $\sigma_3$ is the third Pauli matrix,
breaks the $U(4)$ symmetry down to the $U_{\uparrow}(2) \times U_{\downarrow}(2)$ symmetry.
The generators of the $U_{\uparrow}(2) \times U_{\downarrow}(2)$ subgroup are given by
\begin{eqnarray}
P_{s}\otimes I_4 , \quad P_{s}\otimes i \gamma^3 , \quad
P_{s}\otimes \gamma^5 , \quad P_{s}\otimes \frac{1}{2}[\gamma^3,\gamma^5] ,
\label{U2generators}
\end{eqnarray}
where $\gamma^5\equiv i \gamma^0\gamma^1\gamma^2\gamma^3$ and
$P_{s}=(1+s\sigma_{3})/2$ are the projectors on spin-$\uparrow$ ($s=+1$) and
spin-$\downarrow$ ($s=-1$) states. Note that 
$U_{s}(2)\sim U_{s}(1) \times SU_{s}(2) $, where the first matrix in Eq.~(\ref{U2generators}) is a generator
of the $U_{s}(1)$ and last three matrices are the generators of the $SU_{s}(2) $.

The order parameters that describe the breakdown of the $U(4)$ symmetry are
the same as in Ref.~\cite{Gorbar2008PRB}. The charge densities (and
corresponding chemical potentials), which span the QHF order parameters,
are given by:
\begin{eqnarray}
\label{singlet_mu}
\mu_3:&\quad&
 {\Psi^{\dagger}\sigma^3 \Psi} =  
 \sum_{\kappa=K  ,K^{\prime}} \sum_{a =A ,B}
\left(  \psi_{\kappa a \uparrow}^\dagger\psi_{\kappa a \uparrow}
- \psi_{\kappa a \downarrow}^\dagger\psi_{\kappa a \downarrow}\right),\\
\label{triplet_mu}
\tilde{\mu}_{s}:&\quad&
{\Psi^{\dagger}\gamma^3\gamma^5P_{s}\Psi} =
  \psi_{K  As}^\dagger\psi_{K As}
+ \psi_{K Bs}^\dagger \psi_{K Bs}
- \psi_{K^{\prime} As}^\dagger \psi_{K^{\prime}As}
- \psi_{K^{\prime}Bs}^\dagger \psi_{K^{\prime} Bs}\,,
\end{eqnarray}
where
\begin{equation}
\gamma^3\gamma^5 = \left(\begin{array}{cc} I& 0 \\ 0& -I\end{array}\right)
\label{gamma35}
\end{equation}
is the diagonal valley matrix related to the $SU_s(2)$ symmetry.
The chemical potentials
$\mu_3=(\mu_{\uparrow}-\mu_{\downarrow})/2$ and $\tilde{\mu}_s$ are related to the spin and
valley densities, respectively. The latter describes a charge density imbalance between the two valleys
in the Brillouin zone (or the anomalous magnetic moment in the language of relativistic field theory).

Their MC cousins are connected with charge density wave (CDW) and valley polarized CDW.
The corresponding masses are the Dirac mass and the Haldane mass \cite{Haldane1988PRL},
\begin{eqnarray}
\label{triplet_mass}
\tilde{\Delta}_{s}:&\quad&{\bar{\Psi} P_{s} \Psi} =
  \psi_{K  As}^\dagger\psi_{K As}- \psi_{K Bs}^\dagger \psi_{K Bs}
+ \psi_{K^{\prime} As}^\dagger \psi_{K^{\prime}As}- \psi_{K^{\prime}Bs}^\dagger \psi_{K^{\prime} Bs}\,,\\
\label{singlet_mass}
\Delta_{s}: &\quad&{\bar{\Psi}\gamma^3 \gamma^5 P_{s} \Psi} =
  \psi_{K  As}^\dagger\psi_{K As}- \psi_{K Bs}^\dagger \psi_{K Bs}
- (\,\psi_{K^{\prime} As}^\dagger \psi_{K^{\prime}As}
- \psi_{K^{\prime}Bs}^\dagger \psi_{K^{\prime} Bs}\,)\,,
\end{eqnarray}
 respectively.

While the generation of CDW in Eq.~(\ref{triplet_mass}) breaks spontaneously the
$U(2)_{s}$ [more precisely, its subgroup $SU(2)_{s}$],
the generation of valley polarized CDW in Eq.~(\ref{singlet_mass})
does not break this symmetry.
On the other hand, while the Dirac mass term (\ref{triplet_mass}) is even
under time reversal $\cal{T}$, the Haldane mass term (\ref{singlet_mass})
is $\cal{T}$-odd (for a recent review of the transformation properties of different
mass terms in graphene, see Ref.~\cite{Gusynin2007review}). It is noticeable
that the mass $\Delta$ was first discussed long ago
in connection with inducing the Chern-Simons term in the effective
action of $(2 + 1)$-dimensional relativistic gauge field theories \cite{Niemi1983PRL}.

\section{Quasiparticle propagator}
\label{secIII}

The inverse {\em bare} quasiparticle propagator in the mixed $(\omega,\mathbf{r})$-representation is given by
\begin{equation}
i\,S^{-1}(\omega;\mathbf{r},\mathbf{r^{\prime}})=[(\omega+\mu)\gamma^0 -
\mu_B B  \gamma^0\sigma^3+v_F(\bm{\pi}\cdot\bm{\gamma})]\delta(\mathbf{r}-\mathbf{r}^\prime).
\label{inversebare}
\end{equation}
(To simplify the representation of all formulas, we omit the spin index in this section.)
Similarly, the general structure of the {\em full} fermion propagator for quasiparticles
of a fixed spin has the following form:
\begin{equation}
i\,G^{-1}(\omega;\mathbf{r},\mathbf{r}^\prime)=\left\{\gamma^0 \omega
+v_F \hat{F}^{+} \, (\bm{\pi}\cdot\bm{\gamma})
+\hat{\Sigma}^{+} \right\}\delta(\mathbf{r}-\mathbf{r}^\prime),
\label{inversefull}
\end{equation}
where $\hat{F}^{+}$ and $\hat{\Sigma}^{+}$ can be viewed as generalized wave-function renormalization
and self-energy operators, respectively. In the special case of the bare propagator
in Eq.~(\ref{inversebare}), the corresponding functions are $\hat{F}_{\rm bare}^{+}=1$ and
$\hat{\Sigma}_{\rm bare}^{+}=(\mu-\mu_B B\sigma^3)\gamma^0$.

By definition, $\hat{F}^{+}$ and $\hat{\Sigma}^{+}$ are functions of the three mutually commuting
dimensionless operators: $(\bm{\pi}\cdot\bm{\gamma})^{2}\ell^2$, $\gamma^0$ and
$i s_{\perp}\gamma^1\gamma^2$, where $s_{\perp}={\rm sgn}(eB)$ and 
$\ell=\sqrt{\hbar c/|eB|}$ is the magnetic length. Taking into account that
$(\gamma^0)^2=1$ and $(i s_{\perp}\gamma^1\gamma^2)^2=1$, the operators $\hat{F}^{+}$
and $\hat{\Sigma}^{+}$ can be written in the following form:
\begin{eqnarray}
\hat{F}^{+}&=&  f +\gamma^0g
+i s_{\perp}\gamma^1\gamma^2 \tilde{g}+ i s_{\perp}\gamma^0 \gamma^1\gamma^2 \tilde{f},
\label{def-F+TXT}\\
\hat{\Sigma}^{+}&=& \tilde{\Delta} + \gamma^0\mu
+  i s_{\perp}\gamma^1\gamma^2 \tilde\mu+  i s_{\perp}\gamma^0 \gamma^1\gamma^2 \Delta,
\label{def-S+TXT}
\end{eqnarray}
where $f$, $\tilde{f}$, $g$, $\tilde{g}$, $\tilde{\Delta}$, $\Delta$, $\mu$, and
$\tilde\mu$ are functions of only one operator, $(\bm{\pi}\cdot\bm{\gamma})^{2}\ell^2$.
Note that for the functions $\mu$, $\tilde\mu$ and $\tilde{\Delta}$, $\Delta$ we keep the same
notations as for the parameters $\mu$, $\tilde\mu$ and $\tilde{\Delta}$, $\Delta$ in
Eqs. (\ref{singlet_mu}), (\ref{triplet_mu}) and Eqs. (\ref{triplet_mass}), (\ref{singlet_mass}),
respectively.

It is obvious from the representations  in Eqs.~(\ref{def-F+TXT}) and (\ref{def-S+TXT})
that $\hat{F}^{+}$ and $\hat{\Sigma}^{+}$
do not necessarily commute with $(\bm{\pi}\cdot\bm{\gamma})$. It is convenient, therefore, to introduce
two other functions $\hat{F}^{-}$ and $\hat{\Sigma}^{-}$, which satisfy the relations:
\begin{eqnarray}
\hat{F}^{+} (\bm{\pi}\cdot\bm{\gamma}) &=& (\bm{\pi}\cdot\bm{\gamma}) \hat{F}^{-},
\label{relation-FplusFminus}\\
\hat{\Sigma}^{+} (\bm{\pi}\cdot\bm{\gamma}) &=& (\bm{\pi}\cdot\bm{\gamma}) \hat{\Sigma}^{-}.
\label{relation-SigmaplusSigmaminus}
\end{eqnarray}
As follows from their definition, the explicit representations of these functions read:
\begin{eqnarray}
\hat{F}^{-}&=&  f -\gamma^0g
-i s_{\perp}\gamma^1\gamma^2 \tilde{g}+ i s_{\perp}\gamma^0 \gamma^1\gamma^2 \tilde{f},
\label{def-F-TXT}\\
\hat{\Sigma}^{-}&=& \tilde{\Delta} - \gamma^0\mu
- i s_{\perp}\gamma^1\gamma^2 \tilde\mu+  i s_{\perp}\gamma^0 \gamma^1\gamma^2 \Delta.
\label{def-S-TXT}
\end{eqnarray}
These are obtained from $\hat{F}^{+}$ and $\hat{\Sigma}^{+}$ by reversing the signs in front of the two terms that
anticommute with $ (\bm{\pi}\cdot\bm{\gamma}) $.

The physical meaning of the functions $\tilde{\Delta}$, $\Delta$, $\mu$, and $\tilde\mu$ that appear in the
definition of $\hat{\Sigma}^{\pm}$ is straightforward: $\tilde{\Delta}$ is the Dirac mass function, $\Delta$ is
the Haldane (time-reversal odd) mass function, $\mu$ is the charge density chemical potential, and
$\tilde\mu$ is the chemical potential for charge density imbalance between the two valleys in the Brillouin
zone. As for the functions $f$, $\tilde{f}$, $g$, and $\tilde{g}$ that appear in the definition of $\hat{F}^{\pm}$,
they are various structures in the wave function renormalization operator.

It may appear that, in the most general case, the full propagator (\ref{inversefull}) can also include
another wave-function renormalization, multiplying the frequency term $\gamma^0 \omega$. This is
not the case, however, because this Dirac structure is already included in the self-energy
$\hat{\Sigma}^{+}$, which may depend on $\omega$ in general. We note at the same time that the
solution for $\hat{\Sigma}^{+}$ will turn out to be independent of $\omega$ in the instantaneous
approximation utilized in this study. This fact is also one of the reasons that makes it particularly
convenient to separate the term $\gamma^0 \omega$ from the generalized self-energy operator
$\hat{\Sigma}^{+}$ in (\ref{inversefull}).

As mentioned earlier, functions $f$, $\tilde{f}$, $g$, $\tilde{g}$, $\tilde{\Delta}$, $\Delta$, $\mu$, and
$\tilde\mu$ are functions of $(\bm{\pi}\cdot\bm{\gamma})^{2}\ell^2$, whose eigenvalues are nonpositive
even integers: $-2n\equiv-(2N+1+ s_\perp s_{12})$, where $N=0,1,2,\ldots$ is the orbital quantum
number and $s_{12}=\pm 1$ is the sign of the pseudospin projection. Therefore, in what follows,
it will be convenient to use the following eigenvalues of the operators $\hat{F}^{\pm}$ and $\hat{\Sigma}^{\pm}$
(see Appendix~\ref{appFermionPropagator} for more details),
\begin{eqnarray}
F^{s_0,s_{12} }_{n}&\equiv &f_{n} + s_0g_{n} + s_{12} \tilde{g}_{n}+ s_0 s_{12} \tilde{f}_{n}\,,
\label{F-n}\\
 \Sigma^{s_0,s_{12} }_{n}&\equiv &\tilde{\Delta}_{n} + s_0\mu_{n} + s_{12} \tilde{\mu}_{n}+ s_0 s_{12}
 \Delta_{n}\,,
\label{Sigma-n}
\end{eqnarray}
where $f_{n}$, $\tilde{f}_{n}$, $g_{n}$, $\tilde{g}_{n}$, $\tilde{\Delta}_{n}$, $\Delta_{n}$, $\mu_{n}$,
and $\tilde\mu_{n}$ are the eigenvalues of the corresponding coefficient operators in the $n$th
LL state. Further, $s_0=\pm 1$ and $s_{12}=\pm 1$ are the eigenvalues of $\gamma^0$ and
$is_{\perp}\gamma^1\gamma^2$, respectively.

In terms of eigenvalues, the inverse propagator is derived in Appendix~\ref{appFermionPropagator}.
Its final form reads
\begin{eqnarray}
i\,G^{-1}(\omega;\mathbf{r},\mathbf{r}^\prime) &=& e^{i\Phi(\mathbf{r},\mathbf{r}^{\prime})}
i\,\tilde{G}^{-1}(\omega;\mathbf{r}-\mathbf{r}^\prime)  ,
\label{G-inverse}\\
i\,\tilde{G}^{-1}(\omega;\mathbf{r}) &=&
\frac{e^{-\xi/2}}{2\pi \ell^2}
\sum\limits_{n=0}^{\infty}
\sum_{\sigma=\pm 1}
\sum_{s_0=\pm 1}
\Big\{ s_0 \omega L_{n}(\xi)
+\left[s_0 \mu_{n,\sigma}+\tilde{\Delta}_{n,\sigma}  \right]
\left[\delta^{s_{0}}_{-\sigma} L_{n}(\xi)+\delta^{s_{0}}_{+\sigma} L_{n-1}(\xi) \right]
 \nonumber\\
&&\hspace{2in}
+\frac{i v_F}{\ell^2}(\bm{\gamma}\cdot\mathbf{r})( f_{n,\sigma}-s_0 g_{n,\sigma})L_{n-1}^1(\xi)
\Big\}
{\cal P}_{s_0,s_{0}\sigma},
\label{tildeG-inverse}
\end{eqnarray}
where $L^{\alpha}_n$ are Laguerre polynomials ($L^{0}_n \equiv L_n$). 
We also introduced the following short-hand notations:
\begin{eqnarray}
\xi = \frac{(\mathbf{r}-\mathbf{r}^{\prime})^2}{2\ell^2},\qquad
\Phi(\mathbf{r},\mathbf{r}^{\prime})= -s_{\perp} \frac{(x+x^\prime)(y-y^\prime)}{2\ell^2},
\quad \mbox{(Schwinger phase)}
\end{eqnarray}
and
\begin{eqnarray}
\mu_{n,\sigma} = \mu_{n}+\sigma\tilde\mu_{n},& \quad&
\tilde{\Delta}_{n,\sigma} = \tilde{\Delta}_{n}+\sigma\Delta_{n},
\label{effective-parameters}\\
f_{n,\sigma} = f_{n}+\sigma\tilde{f}_{n},&\quad&
g_{n,\sigma} = g_{n}+\sigma\tilde{g}_{n}.
\end{eqnarray}
Note that, by definition, the Laguerre polynomials $L^{\alpha}_n$ with negative $n$ are identically zero.
Finally, ${\cal P}_{s_0,s_{12}}$ are the projectors in the Dirac space,
\begin{eqnarray}
{\cal P}_{s_0,s_{12}} =\frac{1}{4}(1+s_0\gamma_0)(1+s_{12}i s_{\perp}\gamma^1\gamma^2),
\quad \mbox{with}\quad s_0,s_{12}=\pm 1 .
\label{projDspace}
\end{eqnarray}
Similarly, the expression for the propagator itself reads
\begin{eqnarray}
G(\omega;\mathbf{r},\mathbf{r}^\prime) &=& e^{i\Phi(\mathbf{r},\mathbf{r}^{\prime})}
\tilde{G}(\omega;\mathbf{r}-\mathbf{r}^\prime) ,
\label{G-itself}\\
\tilde{G}(\omega;\mathbf{r}) &=&
i \frac{e^{-\xi/2}}{2\pi \ell^2}
\sum\limits_{n=0}^{\infty}
\sum_{\sigma=\pm 1} \sum_{s_0=\pm 1}
\Big\{
\frac{s_0 (\omega+\mu_{n,\sigma})-\tilde{\Delta}_{n,\sigma}}{(\omega+\mu_{n,\sigma})^2-E_{n,\sigma}^2}
\left[ \delta^{s_{0}}_{-\sigma} L_{n}(\xi)+ \delta^{s_{0}}_{+\sigma} L_{n-1}(\xi)\right]
\nonumber\\
&&\hspace{1in}
+\frac{i v_F}{\ell^2}(\bm{\gamma}\cdot\mathbf{r}) \frac{f_{n,\sigma}-s_0 g_{n,\sigma}}
{(\omega+\mu_{n,\sigma})^2-E_{n,\sigma}^2}L_{n-1}^1(\xi)
\Big\}{\cal P}_{s_0,s_{0}\sigma} ,
\label{propagator-full-expression}
\end{eqnarray}
where the energies in the lowest and higher LLs are
\begin{eqnarray}
E_{0,\sigma} &=& \sigma \, \tilde{\Delta}_{0,\sigma} = \Delta_{0} +\sigma \, \tilde\Delta_{0} ,\\
E_{n,\sigma} &=& \sqrt{2n(v_F^2/\ell^2)\left[f_{n,\sigma}^2-g_{n,\sigma}^2\right]+\tilde{\Delta}_{n,\sigma}^2},
\quad\mbox{for}\quad n\geq 1.
\end{eqnarray}
The corresponding energies of quasiparticles are determined by the location of the poles of propagator
(\ref{propagator-full-expression}), i.e.,
\begin{eqnarray}
\omega_{0,\sigma} &=& -\mu_{0,\sigma}+E_{0,\sigma},
\label{LLL-energy}\\
\omega_{n,\sigma}^{\pm} &=& -\mu_{n,\sigma}\pm E_{n,\sigma},\quad\mbox{for}\quad n\geq 1.
\end{eqnarray}
Let us note that $\sigma=\pm 1$ is the eigenvalue of matrix
$i s_{\perp}\gamma^0\gamma^1\gamma^2\equiv  s_{\perp}\gamma_3\gamma_5$,
which up to the overall sign $s_{\perp}$ is the quantum number associated with the valley.
This follows from the explicit representation in Eq.~(\ref{gamma35}) and from our convention
for the four-component Dirac spinor, whose first two components are associated with valley
$K$ and the last two components with valley $K^\prime$.

\section{Schwinger--Dyson equation}
\label{secIV}

\begin{figure}
\begin{center}
\includegraphics[width=.75\textwidth]{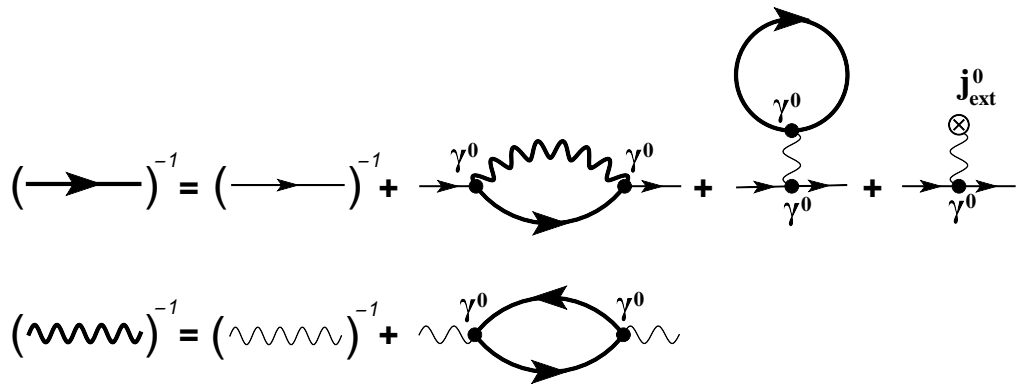}
\caption{The diagrammatic form of the Schwinger--Dyson equations for the electron
and photon propagators in the mean field approximation.}
\label{fig.SD_eq}
\end{center}
\end{figure}

The Schwinger--Dyson (gap) equation for the fermion propagator in the random-phase
approximation (RPA) is
shown diagrammatically in Fig.~\ref{fig.SD_eq}. Note that in contrast to the naive
mean-field approximation, the RPA Coulomb interaction includes the polarization
(screening) effects, which are not negligible in the dynamics responsible for
symmetry breaking in graphene.

It is important to emphasize that the gap equation for the fermion propagator
in Fig.~\ref{fig.SD_eq}
contains two tadpole diagrams. One of them is connected with the Hartree contribution
due to dynamical
charge carriers, while the other takes into account the background charge from the ions in
graphene and in the substrate. The presence of both tadpoles is essential to insure the
overall neutrality of the sample. Indeed, the equation for the gauge field implies
that the two tadpole contributions should exactly cancel that yields
(the Gauss law):
\begin{equation}
j_{\rm ext}^{0}-e\,\mbox{Tr}\left[\gamma^0 G\right]=0\,.
\end{equation}
As clear from the above arguments, this is directly related to the gauge symmetry in the model. 
(Since the Gauss law does not take place in models with contact interactions \cite{Gorbar2008PRB},
there is an analogue of only one tadpole diagram describing the Hartree interaction, which 
contributes to the gap equation.) Thus, the resulting Schwinger--Dyson
equation for the fermion propagator takes the form
\begin{equation}
G^{-1}(t-t^\prime;\mathbf{r},\mathbf{r}^{\prime}) = S^{-1}(t-t^\prime;\mathbf{r},\mathbf{r}^{\prime})
+ e^2\gamma^0\,G(t-t^\prime;\mathbf{r},\mathbf{r}^{\prime})\gamma^0 D(t^\prime-t;\mathbf{r}^{\prime}-\mathbf{r}),
\label{SD}
\end{equation}
where $G(t;\mathbf{r},\mathbf{r}^{\prime})$ is the full fermion propagator and
$D(t;\mathbf{r})$ is the  propagator mediating the Coulomb interaction.

At this point it is instructive to compare the cases of a local four-fermion
interaction and a nonlocal Coulomb interaction. In the case of a local four-fermion
interaction, the right-hand side of the Schwinger--Dyson equation contains
$\delta(\mathbf{r}-\mathbf{r}^{\prime})$ and the fermion propagator only at the point
of coincidence $G(0;\mathbf{r},\mathbf{r})$. This means that
the right-hand side of the Schwinger--Dyson equation is a constant in the momentum space and,
hence, it does not renormalize the kinetic $\bm{\pi} \cdot \bm{\gamma}$ part of the fermion
propagator (i.e., $F^+=1$). Also, in the case of a local four fermion
interaction, $\Sigma^+$ does not depend on the LL index $n$.
Clearly, this simplifies a lot the analysis of the gap equation. The situation changes
in the case of the nonlocal Coulomb interaction.

%%%%%%%%%%%%%%%%%%%%%%%%%%%%%%%%%%%%%%%
%%%%%%%%%%%%%%%  FIGURE (Feynman diagram, SD)  %%%%%%%%%
%%%%%%%%%%%%%%%%%%%%%%%%%%%%%%%%%%%%%%%
%%%%%%%%%%%%%%%%%%%%%%%%%%%%%%%%%%%%%%%
%%%%%%%%%%%%%%%%%%%%%%%%%%%%%%%%%%%%%%%
%%%%%%%%%%%%%%%%%%%%%%%%%%%%%%%%%%%%%%%

Following Ref.~\cite{Gorbar2002PRB}, we consider the instantanenous approximation for the
Coulomb interaction by neglecting the dependence of the photon polarization function
$\Pi(\omega,k)$ on $\omega$. Then, in momentum space, the photon propagator takes
the following form:
\begin{equation}
D(\omega,k)\approx D(0,k)=\frac{i}{\epsilon_0 [k+\Pi(0,k)]} ,
\end{equation}
where $\Pi(0,k)$ is the static polarization function and $\epsilon_0$ is a dielectric constant.
In essence, the instantaneous approximation neglects the retardation of the interaction.
This may be a reasonable approximation for graphene, whose charge carriers propagate
much slower than the speed of light. It should be kept in mind, however, that such an
approximation has a tendency to underestimate the strength of the Coulomb
interaction \cite{Gorbar2002PRB,GGG2010,Gonzalez}.

Unfortunately, it is difficult to find exact solutions of Eq.~(\ref{SD}). Therefore, one has
to use some approximations. Here, we will study the dynamical symmetry breaking in
the model under consideration retaining contributions only of the lowest and several
first LLs. Obviously, this approximation is consistent only if the dynamically
generated gaps are suppressed compared to the Landau scale $\varepsilon_\ell \equiv \sqrt{\hbar v_F^2|eB|/c}$
(it characterizes the energy spectrum of the free theory and is the gap between the lowest and first LLs).

In the instantaneous approximation, the photon propagator reads
\begin{equation}
D(t,\mathbf{r}) = \int\frac{d^2 \mathbf{k}}{(2\pi)^2}  \int\frac{d \omega}{2\pi} D(\omega,k)
e^{-i\omega t+i\mathbf{k}\cdot\mathbf{r}}
= \frac{i}{\epsilon_0} \int_{0}^{\infty} \frac{dk}{2\pi} \frac{k J_{0}(kr)}{k+\Pi(0,k)}
\delta(t)\,.
\label{photon-r}
\end{equation}
By noting that the Schwinger phase on both sides of the gap equation (\ref{SD}) is
the same, we arrive at the following gap equation for the translation invariant part
of the propagator:
\begin{equation}
i\tilde{G}^{-1}(\omega;\mathbf{r}) = i\tilde{S}^{-1}(\omega;\mathbf{r})
-\frac{e^2}{\epsilon_0} \int_{-\infty}^{\infty} \frac{d\Omega}{2\pi} \int_{0}^{\infty} \frac{dk}{2\pi}
\frac{k J_{0}(kr)}{k+\Pi(0,k)} \gamma^0\,\tilde{G}(\Omega;\mathbf{r})\gamma^0\,.
\label{SD-omega-TXT}
\end{equation}
As shown in Appendix~\ref{gap:eqs}, this is equivalent to an infinite set of algebraic equations, see
Eqs.~(\ref{eq-mu0-eff}), (\ref{eq-m0-eff}), (\ref{mu_n:eq}) and (\ref{m_n:eq}).

\section{Numerical results}
\label{secV}

In this section, we present our numerical solutions to the truncated set of gap equations. In contrast to
our previous analysis in Ref.~\cite{Gorbar2008PRB}, where a model with a contact interaction was studied
in detail, here we investigate the effect of the long-range interaction on the dynamics of symmetry breaking
in QH effect in graphene. For this purpose, we will neglect the screening effects,
captured by the polarization function $\Pi(0,k)$. This approximation may be justified in the case
of a strong magnetic field, which is of our main interest. In the case of $\mu=0$, for example, one has 
$\Pi(0,k) \sim \alpha k ( k\ell)$ in infrared \cite{Gorbar2002PRB} and, therefore, the effects of the 
polarization function are suppressed with respect to the bare Coulomb interaction indeed.
This point greatly simplifies the numerical analysis because now the kernel of the gap
equation takes an analytical form. As for a more rigorous study of the dynamics with screening (e.g.,
in the random phase approximation), it will be presented elsewhere \cite{in-preparation}. Note that the
strength of the Coulomb interaction is characterized by the graphene's ``fine structure constant"
$\alpha \equiv {e^2}/{\epsilon_0 v_F}$,
which is approximately equal to $2.2/\epsilon_0$. In our numerical calculations below, we use $\epsilon_0=1$.

In order to use numerical calculations efficiently, it is important to understand all energy scales in the
problem at hand. Ignoring the large energy cutoff due to a finite width of the conductance band, there
are essentially only two characteristic energy scales in the action of graphene: (i) the Landau energy scale
$\varepsilon_\ell = \sqrt{\hbar v_F^2|eB|/c}$ and (ii) the much smaller Zeeman energy
$ Z \equiv  \mu_B B $. In our numerical calculations, we measure all physical quantities
with the units of energy in units of $\varepsilon_\ell $. Numerically, these are
\begin{eqnarray}
\varepsilon_\ell = \sqrt{\hbar v_F^2|eB|/c} =26 \sqrt{B[T]}~\mbox{meV}, \quad
 Z  = \mu_B B =5.8\times 10^{-2}B[T]~\mbox{meV} ,
\end{eqnarray}
where $B[T]$ is the value of the magnetic field measured in Teslas.
The corresponding temperature scales are $\varepsilon_\ell/k_B=300 \sqrt{B[T]}~\mbox{K}$ and
$ Z /k_B=0.67 B[T]~\mbox{K}$.

Note that the magnetic length $\ell=\sqrt{ \hbar c/|eB|} $ and the Landau energy scale $\varepsilon_\ell $ are
related through the Fermi velocity as follows: $\ell={\hbar  v_F}/{\varepsilon_\ell } =  {26~\mbox{nm}}/{\sqrt{B[T]}}$ ,
where we used $v_F/c= 1/300$.

\subsection{Renormalization of the Fermi velocity (weak field)}

Let us start from the simplest analysis, when the role of the dynamical mass parameters
is negligible and there is no significant splitting of the LLs. This is presumably the case when
the magnetic field is not so strong. Even in this case, however, there is a very interesting dynamics
responsible for the renormalization of the Fermi velocity. This is also interesting from experimental
point of view because the renormalized value of the Fermi velocity parameter, which is also a
function of the Landau index $n$, affects the energies of optical transitions \cite{Sadowski,PRL98-197403,Orlita}.

To this end, let us consider only the subset of the gap equations, which involve the wave-function
renormalization $f_n$, see Eq.~(\ref{gap-eq-f_n}).  Even this subset contains an infinite number of gap
equations for each choice of spin, i.e.,
\begin{equation}
f_{n} = 1+ \frac{\alpha}{2} \sum_{n^\prime=1}^{\infty}
\frac{\kappa_{n^\prime-1,n-1}^{(1)} }{n \sqrt{2n^\prime}}
\left[1-n_{F}\left(E_{n^\prime}-\mu \right)
-n_{F}\left(E_{n^\prime}+\mu \right)
\right],
\quad\mbox{for}\quad n\geq1,
\label{gap-eq-fn_TXT}
\end{equation}
where we took into account the definition of the coefficients $\kappa_{n^\prime,n}^{(1)}$ in
Eq.~(\ref{kappa0-Pi0}) and used the approximate expression for the LL energies:
$E_{n,\sigma} \approx E_{n} =\sqrt{2n} \varepsilon_\ell f_{n,\sigma}$, which are independent of
the valley quantum number $\sigma\equiv s_0 s_{12}$. Note that the spin index is also omitted, which
is justified especially at weak fields. Here we use the notation $n_{F}(x)\equiv 1/(e^x+1)$ for the 
Fermi-Dirac distribution function.

Before analyzing the set of equation for the wave-function renormalization $f_n$ numerically,
it is instructive to note that the expression on the right hand side of Eq.~(\ref{gap-eq-fn_TXT})
is formally divergent. Indeed, by taking into account the asymptotes of the kernel coefficients
$\kappa_{n^\prime-1,n-1}^{(1)}$ as $n^\prime\to \infty$, see Eq.~(\ref{kappa-asymptote-1}), we
find that the sum over $n^\prime$ on the right hand side of Eq.~(\ref{gap-eq-fn_TXT}) is
logarithmically divergent. From quantum field theoretical point of view, of course, this is just
an indication that the coupling constant $\alpha$ is also subject to a renormalization \cite{GonzalezPRB1999}.
For our purposes in this study, however, we may simply assume that the sum over the Landau
levels is finite. Indeed, in contrast to actual relativistic models, the effective action for
quasiparticles of graphene is valid only at sufficiently low energies. Moreover, it is also
clear that the energy width of the conducting band of graphene is finite. In fact, it can
be shown that the formal value of the cutoff in the summation over the LL
index $n$ is approximately given by $n_{\rm max}\simeq 10^4/B[T]$ \cite{Roldan}, where $B[T]$ is the
value of the magnetic field in Teslas. In our numerical calculations, we will use a much
smaller cutoff $n_{\rm max}$. For all practical purposes, when dealing with the observables
in lowest few LLs, such a limitation has little effect on the qualitative and in most cases
even quantitative results. Thus, in the rest of this subsection, we choose the value of the cutoff to
be $n_{\rm max}=100$. (We checked that the numerical results for $f_n$ with the cutoffs 
$n_{\rm max}=50$ and $n_{\rm max}=150$ are qualitatively the same. The values of $f_n$ have
a tendency to grow with increasing $n_{\rm max}$. It is understood, of course, that
such a growth should be compensated by the renormalization of the coupling constant in a
more refined approximation.)

%%%%%%%%%%%%%%%%%%%%%%%%%%%%%%%%%%%%%%%
%%%%%%%%%%%%%%%  FIGURE (Feynman diagram, SD)  %%%%%%%%%
%%%%%%%%%%%%%%%%%%%%%%%%%%%%%%%%%%%%%%%
\begin{figure}
\begin{center}
\includegraphics[width=.5\textwidth]{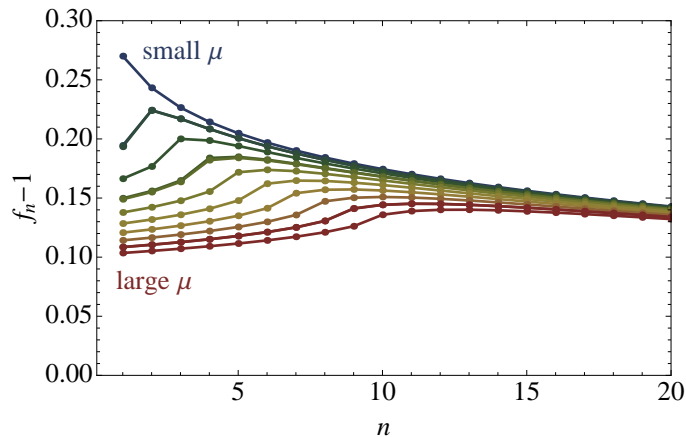}
\caption{The numerical values of the wave function renormalization coefficients $f_n$ vs. 
the LL index $n$ for several fixed values of the chemical potential. }
\label{fig.fn-vs-n}
\end{center}
\end{figure}
%%%%%%%%%%%%%%%%%%%%%%%%%%%%%%%%%%%%%%%
%%%%%%%%%%%%%%%%%%%%%%%%%%%%%%%%%%%%%%%
%%%%%%%%%%%%%%%%%%%%%%%%%%%%%%%%%%%%%%%

The effective Fermi velocity in the $n$th LL is determined by the following relation:
$\tilde{v}_{F,n} = f_n v_{F}$, where the numerical values of the wave function renormalization
are shown in Fig.~\ref{fig.fn-vs-n}. There we show many sets of the results which correspond to
different values of the chemical potentials. The points are the actual data, while the lines
connecting the points are shown for eye guiding the data for fixed values of the chemical
potentials. The data on the top line corresponds to small values of the chemical potential,
$|\mu|<\sqrt{2}\varepsilon_\ell$. The other lines correspond to the chemical potentials in the
energy gaps between $n$th and $(n+1)$th LL (with $n=0,1,2,\ldots$ from top to bottom).
A part of the same data is also given in Table~\ref{tab-fn}. 

In general, the renormalized Fermi velocity 
$\tilde{v}_{F,n}$ is about $10\%$ to $20\%$ larger than its nonrenormalized value $v_{F}$. These
results seem to be somewhat smaller than the predictions in Ref.~\cite{Iyengar}. One should keep 
in mind, however, that there are considerable uncertainties in the theoretical predictions for the
renormalized values of the Fermi velocity. In part, these are associated with a relatively large 
value of the coupling constant in graphene and with the logarithmic running of the wave function 
renormalization itself.

\begin{table}[ht]
\caption{Values of the wave function renormalization $f_{n}$ for several values of the chemical potentials.}
\begin{ruledtabular}
\begin{tabular}{c l l l l l l l}
 & $f_1$ & $f_2$ & $f_3$ & $f_4$ & $f_5$ & $f_6$ & $f_7$ \\
\hline
$|\mu|<\sqrt{2}\varepsilon_\ell$ &  1.270 & 1.243 & 1.227 & 1.214 & 1.205 & 1.197 & 1.190 \\
$\sqrt{2}\varepsilon_\ell<\mu<\sqrt{4}\varepsilon_\ell$ & 1.194 & 1.224 & 1.217 & 1.208 & 1.201 & 1.194 & 1.188 \\
$\sqrt{4}\varepsilon_\ell<\mu<\sqrt{6}\varepsilon_\ell$ & 1.193 & 1.224 & 1.217 & 1.208 & 1.201 & 1.194 & 1.188 \\
$\sqrt{6}\varepsilon_\ell<\mu<\sqrt{8}\varepsilon_\ell$ & 1.166 & 1.177 & 1.200 & 1.199 & 1.194 & 1.189 & 1.184 \\
$\sqrt{8}\varepsilon_\ell<\mu<\sqrt{10}\varepsilon_\ell$ & 1.150 & 1.156 & 1.165 & 1.184 & 1.185 & 1.182 & 1.179 \\
$\sqrt{10}\varepsilon_\ell<\mu<\sqrt{12}\varepsilon_\ell$ &  1.149 & 1.155 & 1.164 & 1.182 & 1.184 & 1.182 & 1.179 \\
$\sqrt{12}\varepsilon_\ell<\mu<\sqrt{14}\varepsilon_\ell$ &  1.138 & 1.142 & 1.148 & 1.156 & 1.172 & 1.174 & 1.173
\label{tab-fn}
\end{tabular}
\end{ruledtabular}
\end{table}

A convenient quantitative measure of the many-particle effects in the transition energies
is given by the prefactors $C_{n,n^\prime}$, which are introduced as deviations from the
non-interacting carriers in graphene,
\begin{equation}
\Delta E_{n,n^\prime} \equiv E_{n^\prime} \pm E_{n} = \left(\sqrt{2n^\prime}\pm \sqrt{2n}\right)
\varepsilon_\ell +\alpha \varepsilon_\ell C_{n,n^\prime},
\label{DeltaE-scaling}
\end{equation}
where, once again, a small Zeeman splitting of the energy levels is ignored.
Note that both terms in (\ref{DeltaE-scaling}) scale as $\sim\sqrt{B}$ and
experimental data on infrared spectroscopy of LLs of graphene clearly confirm
this behavior \cite{Sadowski,PRL98-197403,Orlita}. On the other hand, the dependence 
of the coefficients $C_{n,m}$ on the LL pair allows one to get information on many-body effects.
By making use of our notation for the wave-function renormalization, we obtain
\begin{equation}
C_{n,n^\prime} = \frac{\sqrt{2n^\prime}}{\alpha}(f_{n^\prime}-1)\pm \frac{\sqrt{2n}}{\alpha}(f_{n}-1),
\end{equation}
when $n\neq 0$ and $n^\prime\neq 0$. (For transitions from the $n=0$ level and for transitions to the
$n^\prime=0$ level, the LLL never gives any contribution to the corresponding prefactors.)
By making use of our results for $f_{n}$, we obtain the values of prefactors $C_{n,n^\prime}$. For
transitions between several low-lying LLs, the values of the prefactors are listed in
Table~\ref{table-Cnm}.

\begin{table}[ht]
\caption{Values of some prefactors $C_{n,n^\prime}$, assuming that $\mu=0$.}
\begin{ruledtabular}
\begin{tabular}{c c c c c}
$C_{-1,0}$ & $C_{-1,1}$ & $C_{-2,0}$ & $C_{-2,1}$ & $C_{-2,2}$   \\
\hline
0.176 & 0.352 & 0.224 & 0.400 & 0.448
\end{tabular}
\end{ruledtabular}
\label{table-Cnm}
\end{table}
It is clear that the Coulomb interaction contribution to the LL transitions slightly increases the transition 
energies above their noninteracting values in accordance with experimental data (e.g., see Fig.~3a in 
Ref.~\cite{PRL98-197403}).

\subsection{Lowest Landau level approximation  (strong field)}

In the case of a strong magnetic field, the interaction of charge carriers plays a very
important role. One of the empirical features in this regime is a lifted degeneracy of
the Landau sublevels, associated with spin and valley quantum numbers. This observation
is naturally interpreted as an indication of symmetry breaking in the corresponding ground
states with fixed filling factors. Here we analyze such a dynamics, using
a low-energy effective model with a long-range Coulomb interaction in detail.

To start with the analysis, let us first consider the solutions to the gap equations
for a fixed spin of charge carrier. Keeping only the LLL terms ($n=0$) in the infinite
series on the right hand side of Eqs.~(\ref{eq-mu0-eff}) and
(\ref{eq-m0-eff}), we obtain
\begin{eqnarray}
\mu_{0}^{\rm eff}-\mu&=& \frac{\alpha \varepsilon_\ell }{2} \kappa_{0,0}^{(0)}  \left[
n_{F}\left(\tilde{\Delta}_{0}^{\rm eff}-\mu_{0}^{\rm eff}\right)-n_{F}
\left(\tilde{\Delta}_{0}^{\rm eff}+\mu_{0}^{\rm eff}\right)
\right],\label{eq-mu0-eff-0}\\
\tilde{\Delta}_{0}^{\rm eff}&=& \frac{\alpha \varepsilon_\ell }{2} \kappa_{0,0}^{(0)}    \left[
1-n_{F}\left(\tilde{\Delta}_{0}^{\rm eff}-\mu_{0}^{\rm eff}\right)-n_{F}
\left(\tilde{\Delta}_{0}^{\rm eff}+\mu_{0}^{\rm eff}\right)
\right],
\label{eq-m0-eff-0}
\end{eqnarray}
where we used the notation $\mu^{\rm eff}_0 = \mu_0-\Delta_0$ and $\tilde{\Delta}^{\rm eff}_{0}
=\tilde{\Delta}_{0}-\tilde{\mu}_0$ for the two independent combination of parameters that
determine the spectrum of quasiparticle at the LLL sublevels (see Eq.~(\ref{LLL-energy})),
\begin{eqnarray}
\omega_{0,\downarrow} =-\mu^{\rm eff}_0 - \tilde{\Delta}^{\rm eff}_{0} ,\quad
\omega_{0,\uparrow} = -\mu^{\rm eff}_0 + \tilde{\Delta}^{\rm eff}_{0} .
\end{eqnarray}
At zero temperature, the gap equations become
\begin{eqnarray}
\mu^{\rm eff}_0 &= & \mu+  \frac{\alpha\varepsilon_\ell}{4\sqrt{2\pi}}\sign(\mu^{\rm eff}_0)
\theta(|\mu^{\rm eff}_0|-|\tilde{\Delta}^{\rm eff}_{0}|),\\
|\tilde{\Delta}^{\rm eff}_{0}| &=&  \frac{\alpha\varepsilon_\ell}{4\sqrt{2\pi}}
\theta(|\tilde{\Delta}^{\rm eff}_{0}|-|\mu^{\rm eff}_0|),
\end{eqnarray}
where we took into account the exact value of $\kappa_{0,0}^{(0)} = 1/(2\sqrt{2\pi})$,
see Table~\ref{tab-kappa-0}.
One of the solutions to this set of equations is of the ``magnetic catalysis" type, i.e.,
\begin{eqnarray}
\mu^{\rm eff}_0  &=& \mu ,\quad \mbox{for}\quad |\mu| <  \frac{\alpha\varepsilon_\ell}{4\sqrt{2\pi}}\\
|\tilde{\Delta}^{\rm eff}_{0}| &=&   \frac{\alpha\varepsilon_\ell}{4\sqrt{2\pi}} .
\end{eqnarray}
(Note that, according to this simplified estimate in the LLL approximation, the value of the gap
$|\tilde{\Delta}^{\rm eff}_{0}|$ ranges from $30$~K to $190$~K when the magnetic field changes
from $1$~T to $50$~T, assuming that the coupling constant $\alpha$ is of order $1$.)
The other two solutions are
\begin{eqnarray}
&&-\infty<\mu< \frac{\alpha\varepsilon_\ell}{4\sqrt{2\pi}},\quad
\mu^{\rm eff}_0 =\mu- \frac{\alpha\varepsilon_\ell}{4\sqrt{2\pi}} ,\quad
\tilde{\Delta}^{\rm eff}_{0} = 0 ,\\
&&- \frac{\alpha\varepsilon_\ell}{4\sqrt{2\pi}}<\mu<\infty,\quad
\mu^{\rm eff}_0 =\mu+ \frac{\alpha\varepsilon_\ell}{4\sqrt{2\pi}},\quad
\tilde{\Delta}^{\rm eff}_{0} = 0.
\end{eqnarray}
All of these solutions can be easily found numerically. In addition, the numerical study of the finite
temperature equations shows two extra (probably unstable) solutions, satisfying the approximate condition
$\mu^{\rm eff}_0 \approx  |\tilde{\Delta}^{\rm eff}_{0}| $.  These solutions seem to survive even when one
approaches the limit $T\to 0$. They are lost in the analytical study of the above zero-temperature
equations because they correspond to the vanishing value of the argument in the step-functions.
The ambiguity of the step-function in this case prevents us from finding the same solutions.
Note that all 5 solutions can be continuously continued into each other through a sort of hysteresis
loop without any discontinuities in the values of  $\mu^{\rm eff}_0$ and $|\tilde{\Delta}^{\rm eff}_{0}|$.
This fact alone strongly indicates that some ``intermediate" solutions are metastable or even unstable.

\subsection{Numerical solutions for $\mu$ near the lowest Landau level}

Let us now perform a more realistic numerical analysis by including several LLs and
accounting for the spin degree of freedom in the analysis. The explicit form of the gap equations is presented
in Appendix~\ref{gap:eqs}, see Eqs.~(\ref{eq-mu0-eff}) through (\ref{m_n:eq}). Because of the large number of
the dynamical parameter, in this section, we will use a rather small number for the cutoff $n_{\rm max}=5$ in
the summation over the LLs.

The gap equations for both spins look the same except for
the value of the chemical potential: it is $\mu_{\uparrow} \equiv \mu- Z $ and 
$\mu_{\downarrow} \equiv \mu+ Z $.
By repeating the same analysis as above, we find that there exist many more solutions around the vanishing
value of $\mu$. Keeping only a subset of several qualitatively different solutions with lowest energies, we
find among them a pure Dirac mass solution, 
four types of the Haldane mass solutions and four types of hybrid solutions (see below).
The pure Dirac mass solution has nonzero Dirac masses for both spins and no Haldane masses [i.e., the order 
parameters are triplets with respect to both $SU_{\uparrow}(2)$ and $SU_{\downarrow}(2)$]. Four Haldane 
mass solutions are determined by four possible sign combinations of the two time-reversal breaking masses:
(i) $\Delta_{0,\uparrow}>0$ and $\Delta_{0,\downarrow}>0$;
(ii) $\Delta_{0,\uparrow}>0$ and $\Delta_{0,\downarrow}<0$;
(iii) $\Delta_{0,\uparrow}<0$ and $\Delta_{0,\downarrow}>0$;
(iv) $\Delta_{0,\uparrow}<0$ and $\Delta_{0,\downarrow}<0$.
All of these are characterized by singlet order parameters with respect to both $SU_{\uparrow}(2)$ 
and $SU_{\downarrow}(2)$ symmetry groups. 
Similarly, four hybrid solutions are determined by the following conditions
(i) $\tilde{\Delta}_{0,\uparrow}\neq 0$ and $\Delta_{0,\downarrow}>0$;
(ii) $\tilde{\Delta}_{0,\uparrow}\neq 0$ and $\Delta_{0,\downarrow}<0$;
(iii) $\Delta_{0,\uparrow}>0$ and $\tilde{\Delta}_{0,\downarrow}\neq 0$;
(iv) $\Delta_{0,\uparrow}<0$ and $\tilde{\Delta}_{0,\downarrow}\neq 0$.
The common feature of the hybrid solutions is that one of their order parameters is a triplet with respect to  
$SU_{\uparrow}(2)$ or $SU_{\downarrow}(2)$ group, while the other order parameter is a singlet with respect 
to the other group.

Including many ``intermediate" branches of  solutions, in our numerical analysis we were able to
identify several dozens of non-equivalent solutions in the vicinity of the LLL. The free energies for all 
of these solutions can be easily calculated using Eq.~(\ref{Omega-3}) in Appendix~\ref{C}.

\begin{figure}
\begin{center}
\includegraphics[width=.48\textwidth]{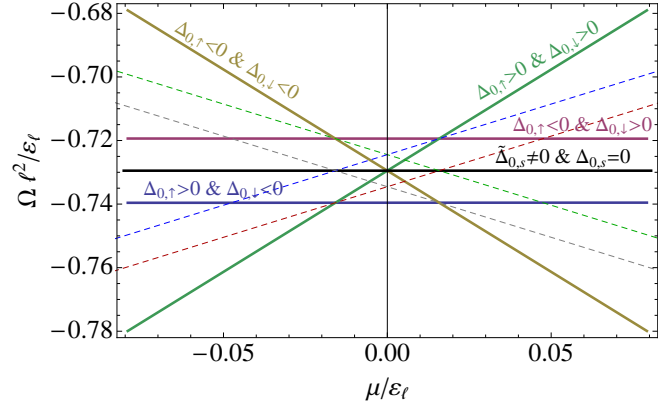}
\caption{Numerical results for free energies of several lowest energy solutions, realized when
the Fermi energy is near $n=0$ LL. The dashed lines correspond to hybrid solutions.}
\label{fig_free-e_LLL_S1S2}
\end{center}
\end{figure}

The free energies of several lowest energy solutions are plotted in Fig.~\ref{fig_free-e_LLL_S1S2}. 
Four singlet type solutions and one triplet solution are shown by solid lines in the figure. Four lowest 
energy hybrid solution are shown by dashed lines.
There are three qualitatively different regions, in which the lowest energy states are different, i.e.,
\begin{eqnarray}
 \mu<- Z : && \Delta_{0,\uparrow}>0~\&~ \Delta_{0,\downarrow}>0,\\
- Z <\mu< Z : &&  \Delta_{0,\uparrow}>0 ~\&~ \Delta_{0,\downarrow}<0,\\
\mu> Z : &&  \Delta_{0,\uparrow}<0 ~\&~ \Delta_{0,\downarrow}<0 .
\end{eqnarray}
At the points $\mu=\pm  Z $ , there also exist hybrid solutions with the same lowest values of the 
energy as the two Haldane mass solutions.

%%%%%%%%%%%%%%%%%%%%%%%%%%%%%%%%%%%%%%%
%%%%%%%%%%%%%%%%%%%%%%%%%%%%%%%%%%%%%%%

\begin{itemize}

\item $\bm{\nu=-2}$ (the LLL is empty). This is a {\em singlet} solution, whose free energy as a function of $\mu$ is shown 
by the green solid line in Fig.~\ref{fig_free-e_LLL_S1S2}.  This solution corresponds to the unbroken
$U_{\uparrow}(2) \times U_{\downarrow}(2)$ symmetry and has the lowest energy for $\mu<- Z $:
\begin{eqnarray}
\tilde{\Delta}_{0,\uparrow}^{\rm eff}=0, \quad & \mu_{0,\uparrow}^{\rm eff}=\mu- Z -\Delta_{0,\uparrow}, &
\quad\Delta_{0,\uparrow}\approx 0.225\varepsilon_{\ell},\\
\tilde{\Delta}_{0,\downarrow}^{\rm eff}= 0, \quad & \mu_{0,\downarrow}^{\rm eff}=\mu+ Z -\Delta_{0,\downarrow},&
\quad\Delta_{0, \downarrow}\approx 0.225\varepsilon_{\ell}.
\end {eqnarray}
In this case the LLL quasiparticle energies are
\begin{eqnarray}
\omega_{0,\uparrow} &=& -\mu+ Z +|\Delta_{0,\uparrow}| >0 ,\qquad (\times 2),\\
\omega_{0,\downarrow} &=& -\mu- Z +|\Delta_{0,\downarrow}|>0,\qquad (\times 2).
\end{eqnarray}
Since none of the LLL sublevels are occupied, this solution corresponds to a $\nu=-2$ state.

\item $\bm{\nu=-1}$ (the LLL is one-quater filled). This is a {\em hybrid} solution, whose free energy as 
a function of $\mu$ is shown by the red dashed line in Fig.~\ref{fig_free-e_LLL_S1S2}. The symmetry in 
the corresponding state is spontaneously broken down to 
$U_{\uparrow}(2) \times U_{\downarrow}^{(K)}(1) \times U_{\downarrow}^{(K^\prime)}(1)$,
where the two latter factors describe $U(1)$ transformations at a fixed spin and a fixed
valley. At $\mu=- Z $, this solution is degenerate in energy with the solutions for the $\nu=-2$ and $\nu=0$ 
cases and is given by
\begin{eqnarray}
\tilde{\Delta}_{0,\uparrow}^{\rm eff}=0, \quad
& \mu_{0,\uparrow}^{\rm eff}=\mu- Z -\Delta_{0,\uparrow},
& \quad \Delta_{0,\uparrow}\approx 0.225\varepsilon_{\ell}, \\
\tilde{\Delta}_{0,\downarrow}^{\rm eff}\approx 0.225\varepsilon_{\ell}, \quad
& \mu_{0,\downarrow}^{\rm eff}=\mu+ Z ,
& \quad \Delta_{0, \downarrow}=0.
\end{eqnarray}
In this case the LLL quasiparticle energies are
\begin{eqnarray}
\omega_{0,\uparrow} &=& -\mu+ Z +|\Delta_{0,\uparrow}| >0 ,\qquad (\times 2),\\
\omega_{0,\downarrow} &=&  -\mu- Z   + \tilde{\Delta}_{0,\downarrow}^{\rm eff} >0,\\
\omega_{0,\downarrow} &=&  -\mu- Z  - \tilde{\Delta}_{0,\downarrow}^{\rm eff} <0.
\end{eqnarray}
Since only one LLL sublevel is occupied, this solution corresponds to a $\nu=-1$ state.

\item $\bm{\nu=0}$ (the half filled LLL; neutral point). This is a {\em singlet} solution. 
Its free energy as a function of $\mu$ is shown by the blue solid line in Fig.~\ref{fig_free-e_LLL_S1S2}.
The symmetry in the corresponding state is
$U_{\uparrow}(2) \times U_{\downarrow}(2)$, but the Zeeman splitting is dynamically
enhanced. This solution has the lowest energy for $- Z <\mu< Z $
\begin{eqnarray}
\tilde{\Delta}_{0,\uparrow}^{\rm eff}=0, \quad &
\mu_{0,\uparrow}^{\rm eff}=\mu- Z -\Delta_{0,\uparrow},
& \quad \Delta_{0,\uparrow}\approx 0.225\varepsilon_{\ell},\\
\tilde{\Delta}_{0,\downarrow}^{\rm eff}= 0, \quad & \mu_{0,\downarrow}^{\rm eff}=\mu+ Z -\Delta_{0,\downarrow},
& \quad \Delta_{0, \downarrow}\approx -0.225\varepsilon_{\ell}.
\end{eqnarray}
In this case the LLL quasiparticle energies are
\begin{eqnarray}
\omega_{0,\uparrow} &=& -\mu+ Z +|\Delta_{0,\uparrow}| >0  ,\qquad (\times 2),\\
\omega_{0,\downarrow} &=& -\mu- Z -|\Delta_{0,\downarrow}| <0,\qquad (\times 2).
\end{eqnarray}
Since two of the LLL  sublevels are occupied, this solution corresponds to a $\nu=0$ state.

\item $\bm{\nu=1}$ (the LLL is three-quater filled). Similarly to the solution for the $\bm{\nu=-1}$ state, 
this is a {\em hybrid} solution. Its free energy as a function of $\mu$ is shown by the gray dashed line 
in Fig.~\ref{fig_free-e_LLL_S1S2}.  The symmetry in the corresponding state is spontaneously broken down to
$U_{\uparrow}^{(K)}(1) \times U_{\uparrow}^{(K^\prime)}(1)\times U_{\downarrow}(2)$.
At $\mu= Z $, this solution is degenerate in energy with the solutions for the $\nu=0$ and $\nu=2$ cases
and is given by
\begin{eqnarray}
\tilde{\Delta}_{0,\uparrow}^{\rm eff}\approx0.225\varepsilon_{\ell},\quad
& \mu_{0,\uparrow}^{\rm eff}=\mu- Z , & \quad \Delta_{i,\uparrow}=0, \\
\tilde{\Delta}_{0,\downarrow}^{\rm eff}= 0,\quad  & \mu_{0,\downarrow}^{\rm eff}=\mu+ Z -\Delta_{0,\downarrow},
& \quad \Delta_{0, \downarrow}\approx-0.225\varepsilon_{\ell}.
\end{eqnarray}
In this case the LLL quasiparticle energies are
\begin{eqnarray}
\omega_{0,\uparrow} &=& -\mu+ Z  + \tilde{\Delta}_{0,\uparrow}^{\rm eff}>0,\\
\omega_{0,\uparrow} &=& -\mu+ Z  - \tilde{\Delta}_{0,\uparrow}^{\rm eff} <0,\\
\omega_{0,\downarrow} &=& -\mu- Z -|\Delta_{0,\downarrow}|<0,\qquad (\times 2).
\end{eqnarray}
Since three of the LLL  sublevels are occupied, this solution corresponds to a $\nu=1$ state.

\item $\bm{\nu=2}$ (the LLL is filled). This is another {\em singlet} solution. 
Its free energy as a function of $\mu$ is shown by the light-brown solid line in Fig.~\ref{fig_free-e_LLL_S1S2}. 
This solution corresponds to the unbroken $U_{\uparrow}(2) \times U_{\downarrow}(2)$ symmetry and has 
the lowest energy for $\mu> Z $:
\begin{eqnarray}
\tilde{\Delta}_{0,\uparrow}^{\rm eff}=0, \quad
& \mu_{0,\uparrow}^{\rm eff}=\mu- Z -\Delta_{0,\uparrow},
& \quad  \Delta_{0,\uparrow}\approx -0.225\varepsilon_{\ell},\\
\tilde{\Delta}_{0,\downarrow}^{\rm eff}= 0, \quad
& \mu_{0,\downarrow}^{\rm eff}=\mu+ Z -\Delta_{0,\downarrow},
& \quad  \Delta_{0, \downarrow}\approx -0.225\varepsilon_{\ell}.
\end{eqnarray}
In this case the LLL quasiparticle energies are
\begin{eqnarray}
\omega_{0,\uparrow} &=& -\mu+ Z -|\Delta_{0,\uparrow}| <0 ,\qquad (\times 2),\\
\omega_{0,\downarrow} &=& -\mu- Z -|\Delta_{0,\downarrow}| <0,\qquad (\times 2).
\end{eqnarray}
Since all LLL  sublevels are occupied, this solution corresponds to a $\nu=2$ state.

\end{itemize}

\subsection{Numerical solutions in the $n=1$ Landau level}

\begin{figure}
\begin{center}
\includegraphics[width=.48\textwidth]{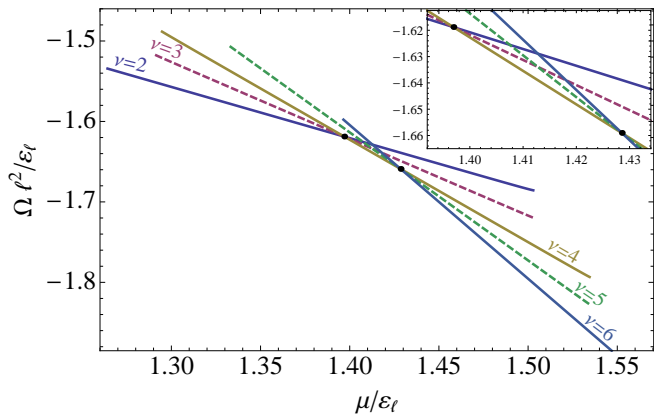}
\caption{Numerical results for the free energies of several lowest energy solutions, realized when
the Fermi energy is near $n=1$ LL.}
\label{fig_free-e_LL1}
\end{center}
\end{figure}

In this subsection, we present the lowest energy solutions in the $n=1$ LL,
when this level is partially or completely filled (see Fig.~\ref{fig_free-e_LL1}).
In this case, we were also able to identify several dozen non-equivalent branches of solutions.
Using Eq.~(\ref{Omega-3}), we selected the ones with the lowest free energies. The main
features of the corresponding solutions are described below. Note that filling of the
$n = 1$ LL leads to changing the dispersion relations also in other LLs.
Therefore, although only parameters $\tilde{\Delta}_1$, $\Delta_1$, $\mu_1$,
and $f_1$ determine the QH plateaus at the $n=1$ LL, we write down their values for
two neighbor levels, the LLL and the $n=2$ LL. (Recall that, according to Eqs.~(\ref{F-n})
and (\ref{Sigma-n}), $\tilde{\Delta}_{n}$, $\Delta_{n}$, $\mu_{n}$, $f_{n}$ are functions
of the LL index $n$.)  Physically, the information concerning the gaps in other 
LLs is relevant for experiments connected with transitions between Landau
levels \cite{PRL98-197403}.

\begin{itemize}

\item $\bm{\nu=3}$ (the $n =1$ LL is one-quater filled). Hybrid solution, which is valid for
$ \mu \simeq \sqrt{2}\varepsilon_\ell- Z $.
The symmetry in this state is spontaneously broken down to 
$U_{\uparrow}(2) \times U_{\downarrow}^{(K)}(1) \times U_{\downarrow}^{(K^\prime)}(1)$:
\begin{eqnarray}
&& \tilde{\Delta}_{0,\uparrow}^{\rm eff}  =0 , \quad
\mu_{0,\uparrow}^{\rm eff} =\mu_{\uparrow}-\Delta_{0,\uparrow} , \quad \Delta_{0,\uparrow} = -0.225\varepsilon_\ell , \\
&& \tilde{\Delta}_{0,\downarrow}^{\rm eff}  =0.052 \varepsilon_\ell , \quad
\mu_{0,\downarrow}^{\rm eff} =\mu_{\downarrow}-\Delta_{0,\downarrow} , \quad \Delta_{0,\downarrow} =-0.277\varepsilon_\ell , \\
&& \tilde{\Delta}_{1,\uparrow}  =0 ,\quad
\mu_{1,\uparrow} =\mu_{\uparrow}+0.053\varepsilon_\ell  ,\quad
\Delta_{1,\uparrow} =-0.067\varepsilon_\ell,\quad
f_{1,\uparrow} = 1.142,\\
&& \tilde{\Delta}_{1,\downarrow}   =-0.018 \varepsilon_\ell , \quad
\mu_{1,\downarrow} =\mu_{\downarrow}+0.148 \varepsilon_\ell,\quad
\Delta_{1,\downarrow} =-0.049\varepsilon_\ell,\quad
f_{1,\downarrow} = 1.103,\\
&& \tilde{\Delta}_{2,\uparrow}  =0 ,\quad
\mu_{2,\uparrow} =\mu_{\uparrow}+0.040\varepsilon_\ell  ,\quad
\Delta_{2,\uparrow} =-0.051\varepsilon_\ell,\quad
f_{2,\uparrow} = 1.111,\\
&& \tilde{\Delta}_{2,\downarrow}   =-0.006 \varepsilon_\ell , \quad
\mu_{2,\downarrow} =\mu_{\downarrow}+0.091 \varepsilon_\ell,\quad
\Delta_{2,\downarrow} =-0.045\varepsilon_\ell,\quad
f_{2,\downarrow} = 1.102.
\end{eqnarray}

\item $\bm{\nu=4}$ (the $n=1$ LL is half filled).
Singlet solution, which is valid for $\sqrt{2}\varepsilon_\ell- Z \lesssim \mu\lesssim \sqrt{2}\varepsilon_\ell+ Z $.
While formally the symmetry of this state is the same as in the action, $U_{\uparrow}(2) \times U_{\downarrow}(2)$, it is characterized by a dynamically
enhanced Zeeman splitting:
\begin{eqnarray}
&& \tilde{\Delta}_{0,\uparrow}^{\rm eff}  =0 , \quad
\mu_{0,\uparrow}^{\rm eff} =\mu_{\uparrow}-\Delta_{0,\uparrow} , \quad \Delta_{0,\uparrow} = -0.225\varepsilon_\ell , \\
&& \tilde{\Delta}_{0,\downarrow}^{\rm eff}  =0 , \quad
\mu_{0,\downarrow}^{\rm eff} =\mu_{\downarrow}-\Delta_{0,\downarrow} , \quad \Delta_{0,\downarrow} = -0.328\varepsilon_\ell , \\
&& \tilde{\Delta}_{1,\uparrow}  =0 ,\quad
\mu_{1,\uparrow} =\mu_{\uparrow}+0.053\varepsilon_\ell  ,\quad
\Delta_{1,\uparrow} =-0.067\varepsilon_\ell,\quad
f_{1,\uparrow} = 1.142,\\
&& \tilde{\Delta}_{1,\downarrow}   =0, \quad
\mu_{1,\downarrow} =\mu_{\downarrow}+0.244 \varepsilon_\ell,\quad
\Delta_{1,\downarrow} =-0.031\varepsilon_\ell,\quad
f_{1,\downarrow} = 1.065\\
&& \tilde{\Delta}_{2,\uparrow}  =0 ,\quad
\mu_{2,\uparrow} =\mu_{\uparrow}+0.040\varepsilon_\ell  ,\quad
\Delta_{2,\uparrow} =-0.051\varepsilon_\ell,\quad
f_{2,\uparrow} = 1.111,\\
&& \tilde{\Delta}_{2,\downarrow}   =0, \quad
\mu_{2,\downarrow} =\mu_{\downarrow}+0.142 \varepsilon_\ell,\quad
\Delta_{2,\downarrow} =-0.039\varepsilon_\ell,\quad
f_{2,\downarrow} = 1.092.
\end{eqnarray}

\item $\bm{\nu=5}$(the $n=1$ LL is three-quater filled).
Hybrid solution, which is valid for $ \mu \simeq \sqrt{2}\varepsilon_\ell+ Z $.
The symmetry in this state is spontaneously broken down to
$U_{\uparrow}^{(K)}(1) \times U_{\uparrow}^{(K^\prime)}(1)\times U_{\downarrow}(2)$:
\begin{eqnarray}
&& \tilde{\Delta}_{0,\uparrow}^{\rm eff}  =0.052 \varepsilon_\ell , \quad
\mu_{0,\uparrow}^{\rm eff} =\mu_{\uparrow}-\Delta_{0,\uparrow} , \quad \Delta_{0,\uparrow} = -0.277\varepsilon_\ell , \\
&& \tilde{\Delta}_{0,\downarrow}^{\rm eff}  =0 , \quad
\mu_{0,\downarrow}^{\rm eff} =\mu_{\downarrow}-\Delta_{0,\downarrow} , \quad \Delta_{0,\downarrow} =-0.328\varepsilon_\ell , \\
&& \tilde{\Delta}_{1,\uparrow}  =-0.018 \varepsilon_\ell  ,\quad
\mu_{1,\uparrow} =\mu_{\uparrow}+0.148\varepsilon_\ell  ,\quad
\Delta_{1,\uparrow} =-0.049\varepsilon_\ell,\quad
f_{1,\uparrow} = 1.103,\\
&& \tilde{\Delta}_{1,\downarrow}   =0, \quad
\mu_{1,\downarrow} =\mu_{\downarrow}+0.244 \varepsilon_\ell,\quad
\Delta_{1,\downarrow} =-0.031\varepsilon_\ell,\quad
f_{1,\downarrow} = 1.065,\\
&& \tilde{\Delta}_{2,\uparrow}  =-0.006 \varepsilon_\ell  ,\quad
\mu_{2,\uparrow} =\mu_{\uparrow}+0.091\varepsilon_\ell  ,\quad
\Delta_{2,\uparrow} =-0.045\varepsilon_\ell,\quad
f_{2,\uparrow} = 1.102,\\
&& \tilde{\Delta}_{2,\downarrow}   =0, \quad
\mu_{2,\downarrow} =\mu_{\downarrow}+0.142 \varepsilon_\ell,\quad
\Delta_{2,\downarrow} =-0.039 \varepsilon_\ell,\quad
f_{2,\downarrow} = 1.092.
\end{eqnarray}

\item $\bm{\nu=6}$ (the $n=1$ LL is filled).
Singlet solution with the unbroken $U_{\uparrow}(2) \times U_{\downarrow}(2)$  symmetry.
It is valid for $\mu\gtrsim \sqrt{2}\varepsilon_\ell+ Z $.that was mentioned in the previous subsection.
\begin{eqnarray}
&& \tilde{\Delta}_{0,\uparrow}^{\rm eff}  =0 , \quad
\mu_{0,\uparrow}^{\rm eff} =\mu_{\uparrow}-\Delta_{0,\uparrow} , \quad \Delta_{0,\uparrow} = -0.328\varepsilon_\ell , \\
&& \tilde{\Delta}_{0,\downarrow}^{\rm eff}  =0 , \quad
\mu_{0,\downarrow}^{\rm eff} =\mu_{\downarrow}-\Delta_{0,\downarrow} , \quad \Delta_{0,\downarrow} = -0.328\varepsilon_\ell , \\
&& \tilde{\Delta}_{1,\uparrow}  =0 ,\quad
\mu_{1,\uparrow} =\mu_{\uparrow}+0.244\varepsilon_\ell  ,\quad
\Delta_{1,\uparrow} =-0.031\varepsilon_\ell,\quad
f_{1,\uparrow} = 1.065,\\
&& \tilde{\Delta}_{1,\downarrow}   =0, \quad
\mu_{1,\downarrow} =\mu_{\downarrow}+0.244 \varepsilon_\ell,\quad
\Delta_{1,\downarrow} =-0.031\varepsilon_\ell,\quad
f_{1,\downarrow} = 1.065,\\
&& \tilde{\Delta}_{2,\uparrow}  =0 ,\quad
\mu_{2,\uparrow} =\mu_{\uparrow}+0.142\varepsilon_\ell  ,\quad
\Delta_{2,\uparrow} =-0.039\varepsilon_\ell,\quad
f_{2,\uparrow} = 1.092,\\
&& \tilde{\Delta}_{2,\downarrow}   =0, \quad
\mu_{2,\downarrow} =\mu_{\downarrow}+0.142 \varepsilon_\ell,\quad
\Delta_{2,\downarrow} =-0.039\varepsilon_\ell,\quad
f_{2,\downarrow} = 1.092.
\end{eqnarray}

\end{itemize}

%%%%%%%%%%%%%%%%%%%%%%%%%%%%%%%%%%%%%%%
%%%%%%%%%%%%%%%%%%%%%%%%%%%%%%%%%%%%%%%

Before concluding this section, it is appropriate to mention that, while all the results for the 
QH states, associated with filling the $n=1$ LLs, are qualitatively similar to those obtained 
in Ref.~\cite{Gorbar2008PRB}, the Coulomb long-range interaction makes all gaps and other 
dynamical parameters functions of the LL index $n$.

\section{Discussion}
\label{Discussion}

The present analysis of integer QH plateaus connected with lifting the fourfold
degeneracy of LLs leads to the following results. The $\nu=0$ QH
state is a QH ferromagnet with significantly dynamically
enhanced Zeeman splitting. The $\nu=1$ QH plateau is described by a CDW 
for the electrons with a spin polarized along the magnetic
field. Such a CDW is absent for the electrons with the
opposite polarization. The solution corresponding to the $\nu=2$ QH plateau
is a singlet and it necessarily includes the Haldane mass.
At the $n=1$ LL, the dynamically generated gaps are much less than those at the
LLL. Therefore, the corresponding QH states could be observed
only in rather strong magnetic fields and very clean samples.

It is noteworthy that the present results obtained in the mean-field approximation with 
the long-range Coulomb interaction are qualitatively similar to those obtained in Ref.~\cite{Gorbar2008PRB},
where a short-range four-fermion interaction was used. The main reasons of this fact are: (a) the
universality of the magnetic catalysis phenomenon, and (b) in both these models, there is
essentially one dimensionfull parameter, the Landau scale $\varepsilon_\ell = \sqrt{\hbar v_F^2|eB|/c}$,
through which the energy spectra of the QH dynamics are expressed. In connection with that, we recall
that the dimensionfull coupling constant $G_{\rm int}$ in the model in Ref.~\cite{Gorbar2008PRB}
is taken as $G_{\rm int} \sim \lambda/\varepsilon_\ell$, and the free dimensionless coupling $\lambda$
plays the role of the Coulomb coupling there. On the other hand, it is noticeable that while
the gaps are constant for the short-range interaction, they decrease with increasing the LL index $n$
in the case of the long-range Coulomb one. Note that without magnetic field energy gaps are functions of a frequency
$\omega$ and a momentum $\bf{k}$ for long-range interactions \cite{Gorbar2002PRB}.
The discrete index $n$ replaces the momentum when magnetic field is switched on.

It should be noted that a recent study of the cyclotron mass in suspended graphene \cite{1104.1396}
suggests a very strong renormalization of the Fermi velocity (at the electron density
$n=10^{9}~\mbox{cm}^{-2}$ it is almost three times larger than the value commonly used).
We note however that the effect of the Fermi velocity renormalization was studied in
several other experiments, such as photoemission spectroscopy \cite{BostwickNaturePh2007},
scanning single-electron transistor measurements of the local electronic
compressibility \cite{MartinNaturePh2008}, and optical infrared measurements in a magnetic 
field \cite{Sadowski,PRL98-197403,Orlita,LiNaturePh2008}.
While the optical measurements observed a departure from the dispersion for noninteracting particles,
its value was small. The present study also suggests that the corresponding renormalization effects
are much smaller than those found in Ref.~\cite{1104.1396}, and, in fact, they are comparable
to those observed in Refs.~\cite{PRL98-197403,AndreiPRB2011}.

It is instructive to compare the present results with the realization of the magnetic
catalysis in $(3 + 1)$-dimensional relativistic theories, such as QED \cite{GusyninPRD1995}
and QCD \cite{MiranskyPRD2002}. A dynamical generation
of only one type of the Dirac mass ($m\equiv \tilde{\Delta}$) is usually considered in relativistic theories in $3 + 1$
dimensions. As we know, the set of the order parameters in graphene is much richer. It is only recently
that new order parameters, analogous to the QHF order parameters and the Haldane
mass, have been studied in QED and QCD \cite{FerrerPRL2009,GorbarPRC2009}. In large part,
the studies of the QHE in graphene have been responsible for those new studies in relativistic
theories. [Note, however, that in $3 + 1$  dimensions the analog of the Haldane mass term describes an
axial-vector current density, rather than a mass \cite{GorbarPRC2009}.]

\begin{acknowledgments}
V.A.M. is grateful to the organizers of the Nobel Symposium on Graphene and 
Quantum Matter for their warm hospitality.
The work of E.V.G and V.P.G. was supported partially by the SCOPES under Grant
No. IZ73Z0-128026 of the Swiss NSF, the Grant No. SIMTECH 246937 of the European
FP7 program, by the SFFR-RFBR Grant ``Application of string theory and field theory
methods to nonlinear phenomena in low dimensional systems",  and by the Program of
Fundamental Research of the Physics and Astronomy Division of the NAS of Ukraine.
The work of V.A.M. was supported by the Natural Sciences and Engineering Research
Council of Canada. The work of I.A.S. is supported in part by a start-up fund from the
Arizona State University and by the U.S. National Science Foundation under Grant
No. PHY-0969844.
\end{acknowledgments}

\appendix

\section{Fermion Green's function}
\label{appFermionPropagator}

\subsection{General structure}

The general structure of the inverse full fermion propagator (for a single spin species)
in a magnetic field is written in Eq.~(\ref{inversefull}). Making use of the two sets of
operators, $\hat{F}^{\pm}$ and $\hat{\Sigma}^{\pm}$, defined in Eqs.(\ref{def-F+TXT}),
(\ref{def-S+TXT}), (\ref{def-F-TXT}), (\ref{def-S-TXT}), we derive the following
formal representation for the full Green's function:
\begin{eqnarray}
G&=& i\left[\gamma^0 \omega +v_F \hat{F}^{+}(\bm{\pi}\cdot\bm{\gamma})+\hat{\Sigma}^{+}) \right]^{-1} \nonumber\\
&=& i\left[\gamma^0 \omega +v_F \hat{F}^{+} (\bm{\pi}\cdot\bm{\gamma})-\hat{\Sigma}^{-} \right]
\left\{  \left[\gamma^0 \omega +v_F \hat{F}^{+} (\bm{\pi}\cdot\bm{\gamma})+\hat{\Sigma}^{+} \right]
\left[\gamma^0 \omega +v_F \hat{F}^{+}(\bm{\pi}\cdot\bm{\gamma})-\hat{\Sigma}^{-} \right]
\right\}^{-1}\nonumber\\
&=& i \left[\gamma^0 \omega +v_F \hat{F}^{+} (\bm{\pi}\cdot\bm{\gamma})-\hat{\Sigma}^{-} \right]
\left\{ \omega^2 +\gamma^0 \omega\left(\hat{\Sigma}^{+}-\hat{\Sigma}^{-}\right)
+v_F^2 \hat{F}^{+} \hat{F}^{-}(\bm{\pi}\cdot\bm{\gamma})^2 -\hat{\Sigma}^{+}\hat{\Sigma}^{-} \right\}^{-1},
\label{operatorId}
\end{eqnarray}
where
\begin{eqnarray}
\gamma^0 \omega( \hat{\Sigma}^{+}-\hat{\Sigma}^{-} )&=& 2 \omega( \mu  + i s_{\perp}
\gamma^0\gamma^1\gamma^2 \tilde\mu), \\
\hat{F}^{+}\hat{F}^{-} &=& (f+i s_{\perp}\gamma^0\gamma^1\gamma^2\tilde{f})^2
- (g+i s_{\perp}\gamma^0\gamma^1\gamma^2\tilde{g})^2 \nonumber\\
&=& f^2+\tilde{f}^2-g^2-\tilde{g}^2 +2 i s_{\perp}\gamma^0 \gamma^1\gamma^2(f\tilde{f}-g\tilde{g}),\\
\hat{\Sigma}^{+}\hat{\Sigma}^{-}&=&  (\tilde{\Delta}+i s_{\perp}\gamma^0\gamma^1\gamma^2 \Delta)^2
- (\mu+i s_{\perp}\gamma^0\gamma^1\gamma^2\tilde{\mu})^2\nonumber\\
&=&\tilde{\Delta}^2+\Delta^2 -\mu^2-\tilde\mu^2
+2 i s_{\perp}\gamma^0 \gamma^1\gamma^2(\tilde{\Delta} \Delta-\mu\tilde\mu),\\
(\bm{\pi}\cdot\bm{\gamma})^2\ell^2 &=& -\bm{\pi}^{2}\ell^2 - (e/c)B i\gamma^1\gamma^2
=-\bm{\pi}^{2}\ell^2 - i s_{\perp}\gamma^1\gamma^2,
\end{eqnarray}
and we recall the definition of the magnetic length $\ell =\sqrt{\hbar c/|eB |}$
and $s_{\perp}\equiv \sign(eB )$. We also used the fact that the operator $\bm{\pi}^{2}\ell^2$
has the eigenvalues $(2N+1)$ with $N=0,1,2,\dots$. In the Landau gauge, $\mathbf{A}=(0,B x)$,
utilized here, the corresponding normalized wave functions read
\begin{eqnarray}
\psi_{Np}(\mathbf{r})=\frac{1}{\sqrt{2\pi \ell}}\frac{1}{\sqrt{2^N N!\sqrt{\pi}}}
H_N\left(\frac{x}{\ell}+p\ell\right)e^{-\frac{1}{2\ell^2}(x+p\ell^2)^2} e^{is_{\perp} py },
\end{eqnarray}
where $H_{N}(x)$ are the Hermite polynomials. These wave functions satisfy the conditions
of normalizability
\begin{eqnarray}
\int d^{2}{r}\psi^{*}_{Np}(\mathbf{r})\psi_{N^{\prime}p^{\prime}}(\mathbf{r})=\delta_{NN^{\prime}}
\delta(p-p^{\prime}),
\end{eqnarray}
and completeness
\begin{eqnarray}
\sum\limits_{N=0}^{\infty}\int\limits_{-\infty}^{\infty} dp\psi_{Np}(\mathbf{r})
\psi^{*}_{Np}(\mathbf{r}^{\prime})
=\delta(\mathbf{r}-\mathbf{r}^{\prime}).
\label{completeness}
\end{eqnarray}
Note that operators $\hat{F}^{\pm}$ and $\hat{\Sigma}^{\pm}$, when acting on eigenstates
$|N,p,s_0,s_{12}\rangle$,
reduce to their eigenvalues:
\begin{eqnarray}
\hat{F}^{\pm}\left[(\bm{\pi}\cdot\bm{\gamma})^2\ell^2,\gamma_0,i s_{\perp}\gamma^1\gamma^2\right]
|N,p,s_0,s_{12}\rangle &=&
\hat{F}^{\pm}\left[-(2N+1+ s_{12}), s_0,s_{12} \right] |N,p,s_0,s_{12}\rangle\nonumber\\
&\equiv & F^{\pm  s_0,\pm s_{12} }_{N+(s_{12}+1)/2}|N,p,s_0,s_{12}\rangle,\\
\hat{\Sigma}^{\pm}\left[(\bm{\pi}\cdot\bm{\gamma})^2\ell^2,\gamma_0,i s_{\perp}\gamma^1\gamma^2\right]
|N,p,s_0,s_{12}\rangle &=&
\hat{\Sigma}^{\pm}\left[-(2N+1+ s_{12}), s_0,s_{12} \right] |N,p,s_0,s_{12}\rangle\nonumber\\
&\equiv & \Sigma^{\pm  s_0,\pm s_{12} }_{N+(s_{12}+1)/2}|N,p,s_0,s_{12}\rangle.
\end{eqnarray}
Taking into account that $s_{12}=\pm 1$, we conclude that the LL index $n = N+(s_{12}+1)/2$
is an nonnegative integer. Therefore, as follows from the definitions in Eqs.~(\ref{def-F+TXT}),
(\ref{def-S+TXT}), (\ref{def-F-TXT}) and (\ref{def-S-TXT}), the eigenvalues are determined
by the following expressions:
\begin{eqnarray}
F^{s_0,s_{12} }_{n}&\equiv &f_{n} + s_0g_{n} + s_{12} \tilde{g}_{n}+ s_0 s_{12} \tilde{f}_{n} , \\
 \Sigma^{s_0,s_{12} }_{n}&\equiv &\tilde{\Delta}_{n} + s_0\mu_{n} + s_{12} \tilde{\mu}_{n}
 + s_0 s_{12} \Delta_{n},
\end{eqnarray}
where $f_{n}$, $\tilde{f}_{n}$, $g_{n}$, $\tilde{g}_{n}$, $\tilde{\Delta}_{n}$,
$\Delta_{n}$, $\mu_{n}$, and $\tilde\mu_{n}$ are the eigenvalues of the
corresponding coefficient operators in the $n$th LL state.

\subsection{Inverse propagator}

Let us introduce the following projectors in the Dirac space:
\begin{eqnarray}
{\cal P}_{s_0,s_{12}} =\frac{1}{4}(1+s_0\gamma_0)(1+s_{12}i s_{\perp}\gamma^1\gamma^2),
\quad \mbox{with}\quad s_0,s_{12}=\pm 1,
\label{proj4}
\end{eqnarray}
Then, we can write the inverse fermion propagator as
\begin{eqnarray}
i\,G^{-1}(\omega;\mathbf{r},\mathbf{r}^\prime) &=& \sum_{s_0,s_{12}=\pm 1}
i\, G_{s_0,s_{12}}^{-1}(\omega;\mathbf{r},\mathbf{r}^\prime){\cal P}_{s_0,s_{12}},
\end{eqnarray}
(note that the projector is multiplied on the right and that the ordering is important)
where
\begin{eqnarray}
&&i\,G_{s_0,s_{12}}^{-1}(\omega;\mathbf{r},\mathbf{r}^\prime)=
\sum\limits_{n=0}^{\infty}\int\limits_{-\infty}^{\infty} dp
\left[s_0 \omega
+v_F (\bm{\pi}\cdot\bm{\gamma})F^{-s_0,-s_{12}}_{n+(s_{12}+1)/2}
+\Sigma^{s_0,s_{12}}_{n+(s_{12}+1)/2} \right] \psi_{np}(\mathbf{r}) \psi_{np}^{*}(\mathbf{r}^{\prime})
\nonumber\\
&=&\frac{e^{-\xi/2+i\Phi(\mathbf{r},\mathbf{r}^{\prime})}}{2\pi \ell^2}\sum\limits_{n=0}^{\infty}
\left[s_0 \omega+\Sigma^{s_0,s_{12}}_{n+(s_{12}+1)/2} \right]
L_{n}(\xi) \nonumber\\
&+& i\frac{e^{-\xi/2+i\Phi(\mathbf{r},\mathbf{r}^{\prime})}}{4\pi \ell^4}\sum\limits_{n=0}^{\infty}
(1+s_{12})v_F F^{-s_0,-1}_{n+1}
\bm{\gamma}\cdot(\mathbf{r}-\mathbf{r}^\prime)L_{n}^1(\xi) \nonumber\\
&+& i\frac{e^{-\xi/2+i\Phi(\mathbf{r},\mathbf{r}^{\prime})}}{4\pi \ell^4}\sum\limits_{n^\prime=0}^{\infty}
(1-s_{12})v_F F^{-s_0,+1}_{n^\prime+1}
\bm{\gamma}\cdot(\mathbf{r}-\mathbf{r}^\prime)L_{n^\prime}^1(\xi)\nonumber\\
&=&\frac{e^{-\xi/2+i\Phi(\mathbf{r},\mathbf{r}^{\prime})}}{2\pi \ell^2}\frac{1+s_{12}}{2}
\sum\limits_{n=0}^{\infty}\left\{\left[s_0 \omega+\Sigma^{s_0,+1}_{n+1} \right]
L_{n}(\xi)+\frac{iv_F}{\ell^2}\bm{\gamma}\cdot(\mathbf{r}-\mathbf{r}^\prime) F^{-s_0,-1}_{n+1}L_{n}^1(\xi)
\right\}\nonumber\\
&+& \frac{e^{-\xi/2+i\Phi(\mathbf{r},\mathbf{r}^{\prime})}}{2\pi \ell^2}\frac{1-s_{12}}{2}
\sum\limits_{n=0}^{\infty}\left\{\left[s_0 \omega+\Sigma^{s_0,-1}_{n} \right]
L_{n}(\xi)+\frac{iv_F}{\ell^2}\bm{\gamma}\cdot(\mathbf{r}-\mathbf{r}^\prime) F^{-s_0,+1}_{n+1}L_{n}^1(\xi)
\right\}\nonumber\\
&=&\frac{e^{-\xi/2+i\Phi(\mathbf{r},\mathbf{r}^{\prime})}}{2\pi \ell^2}
\sum\limits_{n=0}^{\infty}\left\{
\left[s_0 \omega
+ \delta^{s_{12}}_{-1} \Sigma^{s_0,s_{12}}_{n}
+ \delta^{s_{12}}_{+1} \Sigma^{s_0,s_{12}}_{n+1}\right]
L_{n}(\xi)
+\frac{iv_F}{\ell^2}\bm{\gamma}\cdot(\mathbf{r}-\mathbf{r}^\prime)F^{-s_0,-s_{12}}_{n+1}L_{n}^1(\xi)
\right\},
\end{eqnarray}
with $n^\prime=n-1$ (the prime is omitted after the first use)
and the short-hand notation
\begin{eqnarray}
\xi=\frac{(\mathbf{r}-\mathbf{r}^{\prime})^2}{2\ell^2},\qquad
\Phi(\mathbf{r},\mathbf{r}^{\prime})= -s_{\perp} \frac{(x+x^\prime)(y-y^\prime)}{2\ell^2}.
\end{eqnarray}
In the derivation, we changed the integration variable $p\to q \equiv p \ell +
\frac{x+x^\prime}{2\ell}-is_{\perp} \frac{y-y^\prime}{2\ell}$
and took into account that
\begin{eqnarray}
e^{-\frac{1}{2\ell^2}(x+p\ell^2)^2} e^{-\frac{1}{2\ell^2}(x^\prime+p\ell^2)^2}
e^{is_{\perp} p (y - y^\prime) }
&=& e^{-q^2-\xi/2+i\Phi(\mathbf{r},\mathbf{r}^{\prime})},\\
\pi_{x}\psi_{Np} &=& \frac{i}{2\ell}\left[\sqrt{2(N+1)}\psi_{N+1,p} -\sqrt{2N}\psi_{N-1,p} \right],\\
\pi_{y}\psi_{Np} &=& \frac{s_{\perp}}{2\ell}\left[\sqrt{2(N+1)}\psi_{N+1,p} +\sqrt{2N}\psi_{N-1,p} \right],\\
(\bm{\pi}\cdot\bm{\gamma})\psi_{Np} &=&\frac{i}{2\ell}
\left[\sqrt{2(N+1)}(\gamma_1-is_{\perp}\gamma_2)\psi_{N+1,p}
-\sqrt{2N}(\gamma_1+is_{\perp}\gamma_2)\psi_{N-1,p} \right] ,\\
\bm{\pi}^{2}\psi_{Np} &=& \frac{2N+1}{\ell^2}\psi_{Np},
\end{eqnarray}
and integrated over the quantum number $p$ by making use of the formula
$7.378$ in Ref.~\cite{GR},
\begin{equation}
\int\limits_{-\infty}^\infty\,e^{-x^2}H_m(x+y)H_n(x+z)dx
=2^n\pi^{1/2}m!z^{n-m}L_m^{n-m}(-2yz),
\end{equation}
assuming $m\le n$. Here $L^{\alpha}_n$ are the generalized Laguerre polynomials, and
$L_n \equiv L^{0}_n$.

The final representation of the inverse propagator is in the form of the Schwinger phase and
the translation invariant part,
\begin{eqnarray}
i\,G^{-1}(\omega;\mathbf{r},\mathbf{r}^\prime) &=& e^{i\Phi(\mathbf{r},\mathbf{r}^{\prime})}
i\,\tilde{G}^{-1}(\omega;\mathbf{r}-\mathbf{r}^\prime) ,\\
i\,\tilde{G}^{-1}(\omega;\mathbf{r}) &=&
\frac{e^{-\xi/2}}{2\pi \ell^2}
\sum\limits_{n=0}^{\infty}
\sum_{s_0,s_{12}=\pm 1}
\Big\{
\left[s_0 \omega
+ \delta^{s_{12}}_{-1} \Sigma^{s_0,s_{12}}_{n}
+ \delta^{s_{12}}_{+1} \Sigma^{s_0,s_{12}}_{n+1}\right]
L_{n}(\xi) \nonumber\\
&&\hspace{2in}
+\frac{iv_F}{\ell^2}(\bm{\gamma}\cdot\mathbf{r})F^{-s_0,-s_{12}}_{n+1}L_{n}^1(\xi)
\Big\}
{\cal P}_{s_0,s_{12}} ,
\end{eqnarray}
or, it can be written in alternative form as Eq.(\ref{tildeG-inverse}) in the main text.
The inverse of the bare propagator has a similar structure:
\begin{eqnarray}
i\,S^{-1}(\omega;\mathbf{r},\mathbf{r}^\prime) &=& e^{i\Phi(\mathbf{r},\mathbf{r}^{\prime})}
i\,\tilde{S}^{-1}(\omega;\mathbf{r}-\mathbf{r}^\prime) ,\\
i\,\tilde{S}^{-1}(\omega;\mathbf{r}) &=&\frac{e^{-\xi/2}}{2\pi \ell^2}
\sum\limits_{n=0}^{\infty}\sum_{\sigma=\pm 1}\sum_{s_0=\pm 1} \Big\{ s_0 \omega L_{n}(\xi)
+s_0 \mu \left[\delta^{s_{0}}_{-\sigma} L_{n}(\xi)+ \delta^{s_{0}}_{+\sigma} L_{n-1}(\xi) \right]
\nonumber\\
&+&\frac{i v_F}{\ell^2}(\bm{\gamma}\cdot\mathbf{r})L_{n-1}^1(\xi)\Big\}
{\cal P}_{s_0,s_{0}\sigma},
\end{eqnarray}
It should be emphasized that the translation invariant part $\tilde{G}^{-1}(\omega;\mathbf{r})$
is not the inverse of $\tilde{G}(\omega;\mathbf{r})$ used later.

\subsection{Propagator}
\vspace{3mm}

Using the identity in Eq.~(\ref{operatorId}), the spectral expansion of the unit operator
(\ref{completeness}) and the projectors in Eq.~(\ref{proj4}),
we can write the propagator as follows:
\begin{eqnarray}
G(\omega;\mathbf{r},\mathbf{r}^\prime) &=& \sum_{s_0,s_{12}=\pm 1}
G_{s_0,s_{12}}(\omega;\mathbf{r},\mathbf{r}^\prime){\cal P}_{s_0,s_{12}},
\end{eqnarray}
where
\begin{eqnarray}
&& G_{s_0,s_{12}}(\omega;\mathbf{r},\mathbf{r}^\prime)=
i\sum\limits_{n=0}^{\infty}\int\limits_{-\infty}^{\infty} dp
\left[ s_0 \omega +v_F (\bm{\pi}\cdot\bm{\gamma}) F^{-s_0,-s_{12}}_{n+(s_{12}+1)/2}
-\Sigma^{-s_0,-s_{12}}_{n+(s_{12}+1)/2}   \right] \frac{\psi_{np}(\mathbf{r}) \psi_{np}^{*}(\mathbf{r}^{\prime})}
{{\cal M}_{n+(1+s_{12})/2,s_0s_{12}}}
\nonumber\\
&=&i\frac{e^{-\xi/2+i\Phi(\mathbf{r},\mathbf{r}^{\prime})}}{2\pi \ell^2}\sum\limits_{n=0}^{\infty}
\left[ s_0 \omega-\Sigma^{-s_0,-s_{12}}_{n+(s_{12}+1)/2}  \right]
\frac{L_{n}(\xi)}{{\cal M}_{n+(1+s_{12})/2,s_0s_{12}}} \nonumber\\
&-& \frac{e^{-\xi/2+i\Phi(\mathbf{r},\mathbf{r}^{\prime})}}{4\pi \ell^4}\sum\limits_{n=0}^{\infty}
(1+s_{12}) v_F
F^{-s_0,-1}_{n+1}
\bm{\gamma}\cdot(\mathbf{r}-\mathbf{r}^\prime) \frac{L_{n}^1(\xi)}{{\cal M}_{n+1,s_0}} \nonumber\\
&-& \frac{e^{-\xi/2+i\Phi(\mathbf{r},\mathbf{r}^{\prime})}}{4\pi \ell^4}\sum\limits_{n^\prime=0}^{\infty}
(1-s_{12})v_F F^{-s_0,+1}_{n^\prime+1}
\bm{\gamma}\cdot(\mathbf{r}-\mathbf{r}^\prime)\frac{L_{n^\prime}^1(\xi)}{{\cal M}_{n^\prime+1,-s_0}}\nonumber\\
%%%%%%%%%%%%%%%%%%%%%%%%%%%%%%%%%%%%%%%%%
&=&i\frac{e^{-\xi/2+i\Phi(\mathbf{r},\mathbf{r}^{\prime})}}{2\pi \ell^2}\frac{1+s_{12}}{2}
\sum\limits_{n=0}^{\infty}\left\{\left[s_0 \omega-\Sigma^{-s_0,-1}_{n+1}\right]
\frac{L_{n}(\xi)}{{\cal M}_{n+1,s_0}}
+\frac{i v_F}{\ell^2}\bm{\gamma}\cdot(\mathbf{r}-\mathbf{r}^\prime)F^{-s_0,-1}_{n+1}
\frac{L_{n}^1(\xi)}{{\cal M}_{n+1,s_0}}
\right\}\nonumber\\
&+& i\frac{e^{-\xi/2+i\Phi(\mathbf{r},\mathbf{r}^{\prime})}}{2\pi \ell^2}\frac{1-s_{12}}{2}
\sum\limits_{n=0}^{\infty}\left\{\left[s_0 \omega-\Sigma^{-s_0,+1}_{n} \right]
\frac{L_{n}(\xi)}{{\cal M}_{n,-s_0}}
+\frac{i v_F}{\ell^2}\bm{\gamma}\cdot(\mathbf{r}-\mathbf{r}^\prime)F^{-s_0,+1}_{n+1}
\frac{L_{n}^1(\xi)}{{\cal M}_{n+1,-s_0}}
\right\},
\label{Gs0s12}
\end{eqnarray}
where
\begin{eqnarray}
{\cal M}_{n,\sigma} &=& (\omega+\mu_{n,\sigma})^2-E_{n,\sigma}^2, \quad \mbox{with} \quad \sigma=\pm 1, \\
E_{0,\sigma} &=& \sigma \, \tilde{\Delta}_{0,\sigma} = \Delta_{0} +\sigma \, \tilde{\Delta}_{0} ,
\label{E-0-sigma}\\
E_{n,\sigma} &=& \sqrt{2n(v_F^2/\ell^2)\left[f_{n,\sigma}^2-g_{n,\sigma}^2\right]+
\tilde{\Delta}_{n,\sigma}^2},\quad\mbox{for}\quad n\geq 1.
\label{E-n-sigma}
\end{eqnarray}
The energy spectrum at LLL and higher LL's is determined by the location of the poles in Eq.~(\ref{Gs0s12}),
\begin{eqnarray}
\omega_{0,\sigma} &=& -\mu_{0,\sigma}+E_{0,\sigma}, \\
\omega_{n,\sigma}^{\pm} &=& -\mu_{n,\sigma}\pm E_{n,\sigma},\quad\mbox{for}\quad n\geq 1 .
\end{eqnarray}
The final representation of the propagator can be written in the form of the Schwinger phase and
the translation invariant part,
\begin{eqnarray}
G(\omega;\mathbf{r},\mathbf{r}^\prime) &=& e^{i\Phi(\mathbf{r},\mathbf{r}^{\prime})}
\tilde{G}(\omega;\mathbf{r}-\mathbf{r}^\prime) ,\\
\tilde{G}(\omega;\mathbf{r}) &=&
i \frac{e^{-\xi/2}}{2\pi \ell^2}
\sum\limits_{n=0}^{\infty}
\sum_{\sigma=\pm 1} \sum_{s_0=\pm 1}
\Big\{
\frac{s_0 (\omega+\mu_{n,\sigma})-\tilde{\Delta}_{n,\sigma}}{(\omega+\mu_{n,\sigma})^2-E_{n,\sigma}^2}
\left[ \delta^{s_{0}}_{-\sigma} L_{n}(\xi)+ \delta^{s_{0}}_{+\sigma} L_{n-1}(\xi)\right]
\nonumber\\
&&\hspace{1in}
+\frac{i v_F}{\ell^2}(\bm{\gamma}\cdot\mathbf{r}) \frac{f_{n,\sigma}-s_0 g_{n,\sigma}}
{(\omega+\mu_{n,\sigma})^2-E_{n,\sigma}^2}
L_{n-1}^1(\xi)
\Big\}{\cal P}_{s_0,s_{0}\sigma} .
\end{eqnarray}
Note that the free propagator reads
\begin{eqnarray}
S(\omega;\mathbf{r},\mathbf{r}^\prime) &=& e^{i\Phi(\mathbf{r},\mathbf{r}^{\prime})}
\tilde{S}(\omega;\mathbf{r}-\mathbf{r}^\prime) ,\\
\tilde{S}(\omega;\mathbf{r}) &=&
i \frac{e^{-\xi/2}}{2\pi \ell^2}
\sum\limits_{n=0}^{\infty}
\sum_{s_0,s_{12}=\pm 1}
\Big\{
\frac{s_0 (\omega+\mu)\left[ \delta^{s_{12}}_{-1} L_{n}(\xi)+ \delta^{s_{12}}_{+1} L_{n-1}(\xi)\right]}
{(\omega+\mu)^2-2nv_F^2/\ell^2}
+\frac{i v_F}{\ell^2}(\bm{\gamma}\cdot\mathbf{r}) \frac{L_{n-1}^1(\xi)}{(\omega+\mu)^2-2nv_F^2/\ell^2}
\Big\}{\cal P}_{s_0,s_{12}} \nonumber\\
&=&
i \frac{e^{-\xi/2}}{2\pi \ell^2}
\sum\limits_{n=0}^{\infty}
\Big\{ \gamma^0
\frac{(\omega+\mu) \left[ {\cal P}_{-} L_{n}(\xi)+ {\cal P}_{+} L_{n-1}(\xi)\right]}{(\omega+\mu)^2-2nv_F^2/\ell^2}
+\frac{i v_F}{\ell^2}(\bm{\gamma}\cdot\mathbf{r}) \frac{L_{n-1}^1(\xi)}{(\omega+\mu)^2-2nv_F^2/\ell^2}
\Big\} ,
\end{eqnarray}
where $s_{12}\equiv \sigma s_0$ and $ {\cal P}_{\pm}\equiv (1\pm i s_{\perp}\gamma^1\gamma^2)/2$.

\section{Gap equations}
\label{gap:eqs}

The general form of the gap equation is given in Eq.~(\ref{SD}).  By making use
of the photon propagator (\ref{photon-r}),
describing the Coulomb interaction in the instantaneous approximation, and the
mixed $(\omega,\mathbf{r})$-representations
of the fermion propagators, we arrive at the following gap equation:
\begin{equation}
i\tilde{G}^{-1}(\omega;\mathbf{r}) = i\tilde{S}^{-1}(\omega;\mathbf{r})
-\frac{e^2}{\epsilon_0 } \int_{-\infty}^{\infty} \frac{d\Omega}{2\pi} \int_{0}^{\infty} \frac{dk}{2\pi}
\frac{k J_{0}(kr)}{k+\Pi(0,k)} \gamma^0\,
\tilde{G}(\Omega;\mathbf{r})\gamma^0\,.
\label{SD-omega}
\end{equation}
Multiplying both sides of the gap equation (\ref{SD-omega}) by either
$e^{-\xi/2}L_{n}(\xi)$ or $e^{-\xi/2}(\bm{\gamma}\cdot\mathbf{r})L_{n}^1(\xi)$, and then integrating
over $\mathbf{r}$, we find that this is equivalent to following set of equations:
\begin{eqnarray}
\left[\mu_{n,\sigma}-\mu-\sigma \tilde{\Delta}_{n,\sigma}\right] \delta^{s_{0}}_{-\sigma}
 &+& \left[\mu_{n+1,\sigma}-\mu+\sigma \tilde{\Delta}_{n+1,\sigma}\right] \delta^{s_{0}}_{+\sigma}
= -\frac{ie^2}{\epsilon_0} \sum_{n^\prime=0}^{\infty}
 \int_{-\infty}^{\infty} \frac{d\Omega}{2\pi} \int_{0}^{\infty} \frac{dk}{2\pi}
\frac{k {\cal L}_{n^\prime,n}^{(0)}(kl)}{k+\Pi(0,k)}
\nonumber \\
&\times&
\left[
\frac{\Omega+\mu_{n^\prime,\sigma}+\sigma \tilde{\Delta}_{n^\prime,\sigma}}{{\cal M}_{n^\prime,
\sigma}} \delta^{s_{0}}_{-\sigma}
+\frac{\Omega+\mu_{n^\prime+1,\sigma}-\sigma \tilde{\Delta}_{n^\prime+1,\sigma}}{{\cal M}_{n^\prime+1,
\sigma}} \delta^{s_{0}}_{+\sigma}
\right],\nonumber \\
\label{SD-algebraic1}
\\
f_{n,\sigma}+s_0 g_{n,\sigma} -1 &=& \frac{ie^2}{n \epsilon_0} \sum_{n^\prime=1}^{\infty}
 \int_{-\infty}^{\infty} \frac{d\Omega}{2\pi} \int_{0}^{\infty} \frac{dk}{2\pi}
\frac{k {\cal L}_{n^\prime-1,n-1}^{(1)}(kl)}{k+\Pi(0,k)} \frac{f_{n^\prime,\sigma}+s_0 g_{n^\prime,\sigma} }{{\cal M}_{n^\prime,\sigma}},
\quad\mbox{for}\quad n\geq1\,,
\label{SD-algebraic2}
\end{eqnarray}
where
\begin{eqnarray}
{\cal L}^{(0)}_{m,n}=\frac{1}{l^{2}}\int_{0}^{\infty}dr\,r\,e^{-\frac{r^{2}}{2\ell^{2}}}
L_{m}\left(\frac{r^{2}}{2\ell^{2}}\right)L_{n}\left(\frac{r^{2}}{2\ell^{2}}\right)J_{0}(k r)
=(-1)^{m+n}e^{-\frac{k^{2}\ell^{2}}{2}}L_{m}^{n-m}\left(\frac{k^{2}\ell^{2}}{2}\right)
L_{n}^{m-n}\left(\frac{k^{2}\ell^{2}}{2}\right),\label{def-L_mn^0}
\end{eqnarray}
\begin{eqnarray}
{\cal L}^{(1)}_{m,n}=\frac{1}{2l^{4}}\int_{0}^{\infty}dr\,r^{3}\,e^{-\frac{r^{2}}{2\ell^{2}}}
L^{1}_{m}\left(\frac{r^{2}}{2\ell^{2}}\right)L^{1}_{n}\left(\frac{r^{2}}{2\ell^{2}}\right)J_{0}(k r)
=(-1)^{m+n} (m+1) \,e^{-\frac{k^{2}\ell^{2}}{2}}L_{m+1}^{n-m}\left(\frac{k^{2}\ell^{2}}{2}\right)
L_{n}^{m-n}\left(\frac{k^{2}\ell^{2}}{2}\right) \nonumber && \\
=(-1)^{m+n} (n+1) \,e^{-\frac{k^{2}\ell^{2}}{2}}L_{m}^{n-m}\left(\frac{k^{2}\ell^{2}}{2}\right)
L_{n+1}^{m-n}\left(\frac{k^{2}\ell^{2}}{2}\right)\,.&&
\label{def-L_mn^1}
\end{eqnarray}
To obtain the results on the right hand sides, we used the  table integral
 7.422 2 in Ref.~\cite{GR}.

The gap equations can be equivalently rewritten as follows:
\begin{eqnarray}
\mu_{n,\sigma}-\mu-\sigma \tilde{\Delta}_{n,\sigma} &=& -i \alpha \varepsilon_\ell \sum_{n^\prime=0}^{\infty}
\kappa_{n^\prime,n}^{(0)}   \int_{-\infty}^{\infty}  \frac{d\Omega}{2\pi}
\frac{\Omega+\mu_{n^\prime,\sigma}+\sigma \tilde{\Delta}_{n^\prime,\sigma}}{{\cal M}_{n^\prime,\sigma}} ,
\quad\mbox{for}\quad n\geq0, \label{B11} \\
\mu_{n,\sigma}-\mu+\sigma \tilde{\Delta}_{n,\sigma}&=& -i \alpha \varepsilon_\ell \sum_{n^\prime=1}^{\infty}
\kappa_{n^\prime-1,n-1}^{(0)}  \int_{-\infty}^{\infty}  \frac{d\Omega}{2\pi}
\frac{\Omega+\mu_{n^\prime,\sigma}-\sigma \tilde{\Delta}_{n^\prime,\sigma}}{{\cal M}_{n^\prime,\sigma}} ,
\quad\mbox{for}\quad n\geq1,\label{B12}\\
f_{n,\sigma} &=& 1+ i \alpha \varepsilon_\ell \sum_{n^\prime=1}^{\infty}
\frac{\kappa_{n^\prime-1,n-1}^{(1)} }{n}  \int_{-\infty}^{\infty}  \frac{d\Omega}{2\pi}
\frac{f_{n^\prime,\sigma}  }{{\cal M}_{n^\prime,\sigma}},
\quad\mbox{for}\quad n\geq1, \label{B13}\\
g_{n,\sigma}&=& i \alpha \varepsilon_\ell \sum_{n^\prime=1}^{\infty}
\frac{\kappa_{n^\prime-1,n-1}^{(1)} }{n}   \int_{-\infty}^{\infty}  \frac{d\Omega}{2\pi}
 \frac{ g_{n^\prime,\sigma} }{{\cal M}_{n^\prime,\sigma}},
\quad\mbox{for}\quad n\geq1,\label{B14}
\end{eqnarray}
where $\alpha = e^2/(\epsilon_0 v_F)$, $\varepsilon_\ell = v_F/\ell$, and
\begin{eqnarray}
\kappa_{m,n}^{(\rho)} = \int_{0}^{\infty} \frac{dk}{2\pi}
\frac{k \ell  {\cal L}_{m,n}^{(\rho)}(k\ell)}{k+\Pi(0,k)},\quad \rho=0,1.
\end{eqnarray}
When the screening effects are neglected, i.e.,  $\Pi(0,k)=0$, we can use
the explicit form for of ${\cal L}_{m,n}^{(\rho)}$
(with $\rho=0,1$) in Eqs.~(\ref{def-L_mn^0}) and (\ref{def-L_mn^1}) and obtain
the following analytical expressions
for $\kappa_{m,n}^{(\rho)}$ (with $\rho=0,1$):
\begin{eqnarray}
\left. \kappa_{m,n}^{(\rho)}\right|_{\Pi\to0}
&=& \int_{0}^{\infty} \frac{d x }{2\pi} {\cal L}_{mn}^{(\rho)}(x)
=\int\limits_{0}^{\infty} \frac{dx}{2\pi}\int\limits_{0}^{\infty}
dt\,t^{\rho} e^{-t}L^{\rho}_{m}(t)L^{\rho}_{n}(t)J_{0}\left(x\sqrt{2t}\right)
=\frac{1}{2\pi\sqrt{2}}\int\limits_{0}^{\infty}\frac{dt}{t^{1/2}}
t^{\rho} e^{-t}L^{\rho}_{m}(t)L^{\rho}_{n}(t)\nonumber\\
&=&\frac{\Gamma(\rho+1/2)\Gamma(1/2 +n)\Gamma(1/2 +m)}{2\sqrt{2}\pi^{2}m!n!}
{}_{3}F_{2}\left(-m,-n,\rho+1/2;1/2-m,1/2-n;1\right)\nonumber\\
&=&\frac{(-1)^{m+n}}{2\sqrt{2}}\sum\limits_{k=0}^{{\rm min}(m,n)}\frac{\Gamma(\rho+1/2+k)}
{(m-k)!(n-k)!\Gamma(1/2-m+k)\Gamma(1/2-n+k)k!}
\label{kappa0-Pi0} ,
\end{eqnarray}
where we used formula 2.19.14.15 from Ref.~\cite{Laguerre}. The values of
$\kappa_{m,n}^{(\rho)}$ (with $\rho=0,1$) at small values of $m$ and $n$ are given
in Tables~\ref{tab-kappa-0} and \ref{tab-kappa-1}. The leading asymptotes for $n\to \infty$
(at finite $m$) are
\begin{eqnarray}
\left.\kappa_{m,n}^{(0)}\right|_{\Pi\to 0} &\simeq& \frac{1}{2\pi\sqrt{2n}}
+\frac{2m-1}{8\pi(2n)^{3/2}}+O\left(\frac{1}{n^{5/2}}\right)
\quad\mbox{for}\quad n\to\infty ,
\label{kappa-asymptote-0}\\
%%%%%%
\left.\kappa_{m,n}^{(1)}\right|_{\Pi\to 0} &\simeq& \frac{(m+1)}{4\pi\sqrt{2n}}
+\frac{(m+1)(3m-1)}{16\pi(2n)^{3/2}}
+O\left(\frac{1}{n^{5/2}}\right)
\quad\mbox{for}\quad n\to\infty .
\label{kappa-asymptote-1}
\end{eqnarray}

\begin{table}[ht]
\caption{Values of $\kappa_{m,n}^{(0)}$ when the effects of polarization tensor are neglected}
\begin{ruledtabular}
\begin{tabular}{l|llllll }
%\hline
$\kappa_{m,n}^{(0)}$ & $m=0$ & $m=1$ & $m=2$ & $m=3$ & $m=4$ & $m=5$ \\
\hline
$n=0$ & $\frac{1}{2\sqrt{2\pi}}$ & $\frac{1}{4\sqrt{2\pi}}$ & $\frac{3}{16\sqrt{2\pi}}$ & $\frac{5}{32\sqrt{2\pi}}$ & $\frac{35}{256\sqrt{2\pi}}$ & $\frac{63}{512\sqrt{2\pi}}$ \\
$n=1$ & $\frac{1}{4\sqrt{2\pi}}$ & $\frac{3}{8\sqrt{2\pi}}$ & $\frac{7}{32\sqrt{2\pi}}$ & $\frac{11}{64\sqrt{2\pi}}$ & $\frac{75}{512\sqrt{2\pi}}$ & $\frac{133}{1024\sqrt{2\pi}}$ \\
$n=2$ & $\frac{3}{16\sqrt{2\pi}}$ & $\frac{7}{32\sqrt{2\pi}}$ & $\frac{41}{128\sqrt{2\pi}}$ & $\frac{51}{256\sqrt{2\pi}}$ & $\frac{329}{2048\sqrt{2\pi}}$ & $\frac{569}{4096\sqrt{2\pi}}$ \\
$n=3$ & $\frac{5}{32\sqrt{2\pi}}$ & $\frac{11}{64\sqrt{2\pi}}$ & $\frac{51}{256\sqrt{2\pi}}$ & $\frac{147}{512\sqrt{2\pi}}$ & $\frac{759}{4096\sqrt{2\pi}}$ & $\frac{1245}{8192\sqrt{2\pi}}$ \\
$n=4$ & $\frac{35}{256\sqrt{2\pi}}$ & $\frac{75}{512\sqrt{2\pi}}$ & $\frac{329}{2048\sqrt{2\pi}}$ & $\frac{759}{4096\sqrt{2\pi}}$ & $\frac{8649}{32768\sqrt{2\pi}}$ & $\frac{11445}{65536\sqrt{2\pi}}$ \\
$n=5$ & $\frac{63}{512\sqrt{2\pi}}$ & $\frac{133}{1024\sqrt{2\pi}}$ & $\frac{569}{4096\sqrt{2\pi}}$ & $\frac{1245}{8192\sqrt{2\pi}}$ & $\frac{11445}{65536\sqrt{2\pi}}$ & $\frac{32307}{131072\sqrt{2\pi}}$ 
%\hline
\label{tab-kappa-0}
\end{tabular}
\end{ruledtabular}
\end{table}

\begin{table}[ht]
\caption{Values of $\kappa_{m,n}^{(1)}$ when the effects of polarization tensor are neglected}
\begin{ruledtabular}
\begin{tabular}{l|llllll }
%\hline
$\kappa_{m,n}^{(1)}$ & $m=0$ & $m=1$ & $m=2$ & $m=3$ & $m=4$ & $m=5$ \\
\hline
$n=0$ & $\frac{1}{4\sqrt{2\pi}}$ & $\frac{1}{8\sqrt{2\pi}}$ & $\frac{3}{32\sqrt{2\pi}}$ & $\frac{5}{64\sqrt{2\pi}}$ & $\frac{35}{512\sqrt{2\pi}}$ & $\frac{63}{1024\sqrt{2\pi}}$ \\
$n=1$ & $\frac{1}{8\sqrt{2\pi}}$ & $\frac{7}{16\sqrt{2\pi}}$ & $\frac{15}{64\sqrt{2\pi}}$ & $\frac{23}{128\sqrt{2\pi}}$ & $\frac{155}{1024\sqrt{2\pi}}$ & $\frac{273}{2048\sqrt{2\pi}}$ \\
$n=2$ & $\frac{3}{32\sqrt{2\pi}}$ & $\frac{15}{64\sqrt{2\pi}}$ & $\frac{153}{256\sqrt{2\pi}}$ & $\frac{171}{512\sqrt{2\pi}}$ & $\frac{1065}{4096\sqrt{2\pi}}$ & $\frac{1809}{8192\sqrt{2\pi}}$ \\
$n=3$ & $\frac{5}{64\sqrt{2\pi}}$ & $\frac{23}{128\sqrt{2\pi}}$ & $\frac{171}{512\sqrt{2\pi}}$ & $\frac{759}{1024\sqrt{2\pi}}$ & $\frac{3495}{8192\sqrt{2\pi}}$ & $\frac{5505}{16384\sqrt{2\pi}}$ \\
$n=4$ & $\frac{35}{512\sqrt{2\pi}}$ & $\frac{155}{1024\sqrt{2\pi}}$ & $\frac{1065}{4096\sqrt{2\pi}}$ & $\frac{3495}{8192\sqrt{2\pi}}$ & $\frac{57225}{65536\sqrt{2\pi}}$ & $\frac{67365}{131072\sqrt{2\pi}}$ \\
$n=5$ & $\frac{63}{1024\sqrt{2\pi}}$ & $\frac{273}{2048\sqrt{2\pi}}$ & $\frac{1809}{8192\sqrt{2\pi}}$ & $\frac{5505}{16384\sqrt{2\pi}}$ & $\frac{67365}{131072\sqrt{2\pi}}$ & $\frac{261207}{262144\sqrt{2\pi}}$ 
\label{tab-kappa-1}
%\hline
\end{tabular}
\end{ruledtabular}
\end{table}

The zero temperature gap equations (\ref{B11}) through (\ref{B14}) are straightforwardly generalized to the case
of non-zero temperature by making the replacement $\Omega\to i\Omega_{m}\equiv i \pi T (2m+1)$ and using
the Matsubara sums instead of the frequency
integrations,
\begin{eqnarray}
\int\frac{d\Omega}{2\pi}(\ldots) \to i T\sum_{m=-\infty}^{\infty} (\ldots) \, .
\end{eqnarray}
Then, we use the following table sums:
\begin{eqnarray}
T\sum_{m=-\infty}^{\infty}\frac{1}{(\Omega_{m}-i\mu)^2+a^2}
=\frac{1-n_{F}(a+\mu)-n_{F}(a-\mu)}{2a},\\
T\sum_{m=-\infty}^{\infty}\frac{i\Omega_{m}+\mu}{(\Omega_{m}-i\mu)^2+a^2}
=\frac{n_{F}(a+\mu)-n_{F}(a-\mu)}{2},
\end{eqnarray}
and derive the finite temperature gap equations,
\begin{eqnarray}
\mu_{n,\sigma}-\mu-\sigma \tilde{\Delta}_{n,\sigma} &=&
\frac{\alpha \varepsilon_\ell }{2}\sum_{n^\prime=0}^{\infty}
\kappa_{n^\prime,n}^{(0)}
\Big\{
n_{F}\left(E_{n^\prime,\sigma}-\mu_{n^\prime,\sigma}\right)
-n_{F}\left(E_{n^\prime,\sigma}+\mu_{n^\prime,\sigma}\right)\nonumber\\
&&-\frac{\sigma \tilde{\Delta}_{n^\prime,\sigma}}{E_{n^\prime,\sigma}}
\left[1-n_{F}\left(E_{n^\prime,\sigma}-\mu_{n^\prime,\sigma}\right)
-n_{F}\left(E_{n^\prime,\sigma}+\mu_{n^\prime,\sigma}\right)
\right]
\Big\} ,
\quad\mbox{for}\quad n\geq0,
\label{gap-eq-mu_n-0}\\
\mu_{n,\sigma}-\mu+\sigma \tilde{\Delta}_{n,\sigma}&=&
\frac{\alpha \varepsilon_\ell}{2} \sum_{n^\prime=1}^{\infty}
\kappa_{n^\prime-1,n-1}^{(0)}
\Big\{
n_{F}\left(E_{n^\prime,\sigma}-\mu_{n^\prime,\sigma}\right)
-n_{F}\left(E_{n^\prime,\sigma}+\mu_{n^\prime,\sigma}\right)\nonumber\\
&&+\frac{\sigma \tilde{\Delta}_{n^\prime,\sigma}}{E_{n^\prime,\sigma}}
\left[1-n_{F}\left(E_{n^\prime,\sigma}-\mu_{n^\prime,\sigma}\right)
-n_{F}\left(E_{n^\prime,\sigma}+\mu_{n^\prime,\sigma}\right)
\right]
\Big\} ,
\quad\mbox{for}\quad n\geq1,
\label{gap-eq-mu_n-1}\\
f_{n,\sigma} &=& 1+ \frac{\alpha \varepsilon_\ell}{2} \sum_{n^\prime=1}^{\infty}
\frac{\kappa_{n^\prime-1,n-1}^{(1)} }{n}
\frac{f_{n^\prime,\sigma}  }{E_{n^\prime,\sigma}}
\left[1-n_{F}\left(E_{n^\prime,\sigma}-\mu_{n^\prime,\sigma}\right)
-n_{F}\left(E_{n^\prime,\sigma}+\mu_{n^\prime,\sigma}\right)
\right],
\,\mbox{for}\, n\geq1,
\label{gap-eq-f_n}\\
g_{n,\sigma}&=& \frac{\alpha \varepsilon_\ell}{2} \sum_{n^\prime=1}^{\infty}
\frac{\kappa_{n^\prime-1,n-1}^{(1)} }{n}
 \frac{ g_{n^\prime,\sigma} }{E_{n^\prime,\sigma}}
 \left[1-n_{F}\left(E_{n^\prime,\sigma}-\mu_{n^\prime,\sigma}\right)
-n_{F}\left(E_{n^\prime,\sigma}+\mu_{n^\prime,\sigma}\right)
\right],
\quad\mbox{for}\,\, n\geq1 .
\label{gap-eq-g_n}
\end{eqnarray}
Note that the $n^\prime=0$ term in the first equation can be rewritten in a simpler form:
\begin{eqnarray}
\frac{\alpha \varepsilon_\ell }{2} \kappa_{0,n}^{(0)}
\left[2n_{F}\left(\sigma \tilde{\Delta}_{0,\sigma}-\mu_{0,\sigma}\right)-1\right]
\equiv \frac{\alpha \varepsilon_\ell }{2} \kappa_{0,n}^{(0)}
\tan\left(\frac{\mu_{0}^{\rm eff}-\sigma \tilde{\Delta}_{0}^{\rm eff}}{2T}\right) .
\end{eqnarray}
(As defined in Eq.~(\ref{E-0-sigma}), $E_{0,\sigma}=\sigma \tilde{\Delta}_{0,\sigma}$.)
Separating the LLL from the higher LLs, we derive
\begin{eqnarray}
\mu_{0}^{\rm eff}-\mu-\sigma \tilde{\Delta}_{0}^{\rm eff} &=&
\frac{\alpha \varepsilon_\ell }{2} \kappa_{0,0}^{(0)}
\tan\left(\frac{\mu_{0}^{\rm eff}-\sigma \tilde{\Delta}_{0}^{\rm eff}}{2T}\right)
+\frac{\alpha \varepsilon_\ell }{2}\sum_{n^\prime=1}^{\infty}
\kappa_{n^\prime,0}^{(0)}
\Big\{
n_{F}\left(E_{n^\prime,\sigma}-\mu_{n^\prime,\sigma}\right)
-n_{F}\left(E_{n^\prime,\sigma}+\mu_{n^\prime,\sigma}\right)\nonumber\\
&&-\frac{\sigma \tilde{\Delta}_{n^\prime,\sigma}}{E_{n^\prime,\sigma}}
\left[1-n_{F}\left(E_{n^\prime,\sigma}-\mu_{n^\prime,\sigma}\right)
-n_{F}\left(E_{n^\prime,\sigma}+\mu_{n^\prime,\sigma}\right)
\right]
\Big\},
\end{eqnarray}
which is equivalent to the following set of gap equations for the LLL parameters:
\begin{eqnarray}
\mu_{0}^{\rm eff}-\mu&=& \frac{\alpha \varepsilon_\ell }{2} \kappa_{0,0}^{(0)}  \left[
n_{F}\left(\tilde{\Delta}_{0}^{\rm eff}-\mu_{0}^{\rm eff}\right)-n_{F}\left(\tilde{\Delta}_{0}^{\rm eff}
+\mu_{0}^{\rm eff}\right)\right]\nonumber\\
&+& \frac{\alpha \varepsilon_\ell }{4} \sum_{n^\prime=1}^{\infty} \kappa_{n^\prime,0}^{(0)}
 \left[ n_{F}\left(E_{n^\prime,+}-\mu_{n^\prime,+}\right)-n_{F}\left(E_{n^\prime,+}+\mu_{n^\prime,+}\right)
 +n_{F}\left(E_{n^\prime,-}-\mu_{n^\prime,-}\right)-n_{F}\left(E_{n^\prime,-}+\mu_{n^\prime,-}\right)
\right]\nonumber\\
&- & \frac{\alpha \varepsilon_\ell }{4} \sum_{n^\prime=1}^{\infty} \kappa_{n^\prime,0}^{(0)}
\frac{\tilde{\Delta}_{n^\prime,+}}{E_{n^\prime,+}}
 \left[ 1- n_{F}\left(E_{n^\prime,+}-\mu_{n^\prime,+}\right)-n_{F}\left(E_{n^\prime,+}+\mu_{n^\prime,+}\right)
\right]\nonumber\\
&+& \frac{\alpha \varepsilon_\ell }{4} \sum_{n^\prime=1}^{\infty} \kappa_{n^\prime,0}^{(0)}
\frac{\tilde{\Delta}_{n^\prime,-}}{E_{n^\prime,-}}
 \left[ 1- n_{F}\left(E_{n^\prime,-}-\mu_{n^\prime,-}\right)-n_{F}\left(E_{n^\prime,-}+\mu_{n^\prime,-}\right)
\right],\label{eq-mu0-eff}
\end{eqnarray}
%%%%%%%%%%%%%%%%%
\begin{eqnarray}
\tilde{\Delta}_{0}^{\rm eff}&=& \frac{\alpha \varepsilon_\ell }{2} \kappa_{0,0}^{(0)}    \left[
1-n_{F}\left(\tilde{\Delta}_{0}^{\rm eff}-\mu_{0}^{\rm eff}\right)-n_{F}\left(\tilde{\Delta}_{0}^{\rm eff}
+\mu_{0}^{\rm eff}\right)\right]\nonumber\\
&+& \frac{\alpha \varepsilon_\ell }{4} \sum_{n^\prime=1}^{\infty} \kappa_{n^\prime,0}^{(0)}
 \left[ n_{F}\left(E_{n^\prime,-}-\mu_{n^\prime,-}\right)-n_{F}\left(E_{n^\prime,-}+\mu_{n^\prime,-}\right)
-n_{F}\left(E_{n^\prime,+}-\mu_{n^\prime,+}\right)+n_{F}\left(E_{n^\prime,+}+\mu_{n^\prime,+}\right)
\right]\nonumber\\
&+& \frac{\alpha \varepsilon_\ell }{4} \sum_{n^\prime=1}^{\infty} \kappa_{n^\prime,0}^{(0)}
\frac{\tilde{\Delta}_{n^\prime,+}}{E_{n^\prime,+}}
 \left[ 1- n_{F}\left(E_{n^\prime,+}-\mu_{n^\prime,+}\right)-n_{F}\left(E_{n^\prime,+}+\mu_{n^\prime,+}\right)
\right]\nonumber\\
&+& \frac{\alpha \varepsilon_\ell }{4} \sum_{n^\prime=1}^{\infty} \kappa_{n^\prime,0}^{(0)}
\frac{\tilde{\Delta}_{n^\prime,-}}{E_{n^\prime,-}}
 \left[ 1- n_{F}\left(E_{n^\prime,-}-\mu_{n^\prime,-}\right)-n_{F}\left(E_{n^\prime,-}+\mu_{n^\prime,-}\right)
\right],
\label{eq-m0-eff}
\end{eqnarray}
where the two independent combinations of the LLL parameters in the gap equations are
\begin{eqnarray}
\mu^{\rm eff}_0 =  \mu_0-\Delta_0,\quad
\tilde{\Delta}^{\rm eff}_{0} =\tilde{\Delta}_{0}-\tilde{\mu}_0 .
\end{eqnarray}
It is appropriate to mention that the parameters $\mu_0$ and $\Delta_0$ cannot be unambiguously defined,
while their combination $\mu^{\rm eff}_0$ can be and it is the only combination with a well defined physical
meaning. The same is true for $\tilde{\Delta}_{0}$ and $\tilde{\mu}_0$, which are related to physical
observables only through the effective Dirac mass $\tilde{\Delta}^{\rm eff}_{0} $.

The remaining equations for the chemical potentials and masses in the higher ($n\geq1$) LLs read
\begin{eqnarray}
\mu_{n,\sigma} &=& \mu+\frac{\alpha \varepsilon_\ell }{2} \kappa_{0,n}^{(0)}
\left[2n_{F}\left(\sigma \tilde{\Delta}_{0}^{\rm eff}-\mu_{0}^{\rm eff}\right)-1\right]
\nonumber\\
&+&\frac{\alpha \varepsilon_\ell }{2}\sum_{n^\prime=1}^{\infty}
\frac{\kappa_{n^\prime,n}^{(0)}  + \kappa_{n^\prime-1,n-1}^{(0)}  }{2}
\left[
n_{F}\left(E_{n^\prime,\sigma}-\mu_{n^\prime,\sigma}\right)
-n_{F}\left(E_{n^\prime,\sigma}+\mu_{n^\prime,\sigma}\right)\right]
\nonumber\\
&-&\sigma \frac{\alpha \varepsilon_\ell }{2}\sum_{n^\prime=1}^{\infty}
\frac{\kappa_{n^\prime,n}^{(0)}  - \kappa_{n^\prime-1,n-1}^{(0)}  }{2}
\frac{ \tilde{\Delta}_{n^\prime,\sigma}}{E_{n^\prime,\sigma}}
\left[1-n_{F}\left(E_{n^\prime,\sigma}-\mu_{n^\prime,\sigma}\right)
-n_{F}\left(E_{n^\prime,\sigma}+\mu_{n^\prime,\sigma}\right)
\right] ,\label{mu_n:eq}\\
%%%%%%%%
\tilde{\Delta}_{n,\sigma}&=& \sigma\frac{\alpha \varepsilon_\ell }{2} \kappa_{0,n}^{(0)}
\left[1-2n_{F}\left(\sigma \tilde{\Delta}_{0}^{\rm eff}-\mu_{0}^{\rm eff}\right)\right]
\nonumber\\
&-&\sigma\frac{\alpha \varepsilon_\ell }{2}\sum_{n^\prime=1}^{\infty}
\frac{\kappa_{n^\prime,n}^{(0)} - \kappa_{n^\prime-1,n-1}^{(0)}  }{2}
\left[
n_{F}\left(E_{n^\prime,\sigma}-\mu_{n^\prime,\sigma}\right)
-n_{F}\left(E_{n^\prime,\sigma}+\mu_{n^\prime,\sigma}\right)\right]
\nonumber\\
&+&\frac{\alpha \varepsilon_\ell }{2}\sum_{n^\prime=1}^{\infty}
\frac{\kappa_{n^\prime,n}^{(0)}  + \kappa_{n^\prime-1,n-1}^{(0)}  }{2}
\frac{ \tilde{\Delta}_{n^\prime,\sigma}}{E_{n^\prime,\sigma}}
\left[1-n_{F}\left(E_{n^\prime,\sigma}-\mu_{n^\prime,\sigma}\right)
-n_{F}\left(E_{n^\prime,\sigma}+\mu_{n^\prime,\sigma}\right)
\right] .
\label{m_n:eq}
\end{eqnarray}

\section{Free Energy Density}
\label{C}

\begin{figure}
\begin{center}
\includegraphics[width=.75\textwidth]{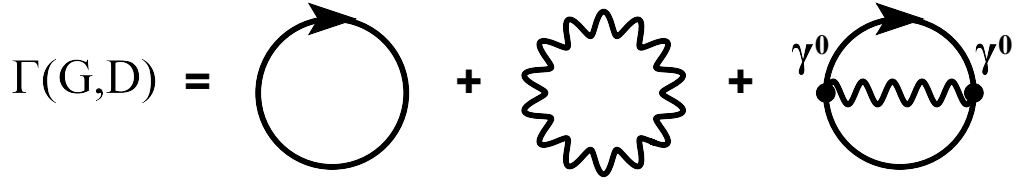}
\caption{The diagrammatic form for the effective action at two loops.}
\label{fig.effective-action}
\end{center}
\end{figure}

Here we calculate the free energy density $\Omega$ in monolayer graphene with dynamically
generated self-energy corrections and wave function renormalization by following the same
approach, based on the Baym-Kadanoff formalism \cite{potential}, that was used in
Ref.~\cite{Gorbar2008PRB} in the case of a model with a contact four-fermion interaction.
In the model with the Coulomb interaction, studied in this paper, the corresponding effective
action $\Gamma$ takes the form:
\begin{eqnarray}
\Gamma(G,D) &=&  {-i}\,\mbox{Tr}\left[\mbox{Ln} G^{-1} +S^{-1}G-1\right]
+\frac{i}{2}\left[\mbox{Ln} D^{-1} +D_{0}^{-1}D-1\right] \nonumber\\
&-&\frac{ie^2}{2}\int d^3u \int d^3 u^\prime \mbox{tr}\left[\gamma^0 G(u,u^\prime)\gamma^0 G(u^\prime,u)\right]
D(u-u^\prime),
\label{Gamma-general}
\label{potential}
\end{eqnarray}
where $u=(t,\mathbf{r})$. The trace, the logarithm, and the product $S^{-1}G$ are taken in the functional 
sense, and $G = \mbox{diag}(G_{\uparrow}, G_{\downarrow})$. The diagrammatic form of this effective
action is shown  in Fig.~\ref{fig.effective-action}. It is instructive to compare this action
with the analogous action in the model with a contact four-fermion interaction \cite{Gorbar2008PRB}.
In contrast to that model, here we have only one type of a two-loop diagram. It is responsible for
the exchange interaction. The other ``dumbbell" diagram (i.e., two fermion loops connected by
the free photon propagator), which would be responsible for the direct (Hartree) interaction, does
not appear in the present model with the gauge interaction. The absence of such a diagram is a
formal consequence of the Gauss's neutrality condition,
\begin{equation}
j_{\rm ext}^{0}(u)-e\,\mbox{tr}\left[\gamma^0 G(u,u)\right]=0,
\end{equation}
where $j_{\rm ext}^{0}(u)$ is the charge density of the substrate and carbon ions that compensate
the overall nonzero charge density of free carriers in the graphene monolayer.

%%%%%%%%%%%%%%%%%%%%%%%%%%%%%%%%%%%%%%%
%%%%%%%%%%%%%%%  FIGURE (free energy)  %%%%%%%%%
%%%%%%%%%%%%%%%%%%%%%%%%%%%%%%%%%%%%%%%
%%%%%%%%%%%%%%%%%%%%%%%%%%%%%%%%%%%%%%%
%%%%%%%%%%%%%%%%%%%%%%%%%%%%%%%%%%%%%%%
%%%%%%%%%%%%%%%%%%%%%%%%%%%%%%%%%%%%%%%

The free energy density $\Omega$ is expressed through $\Gamma$ as follows: $\Omega = -\Gamma/TV$,
where $TV$ is a space-time volume. When the fermion propagator satisfies the gap equation (\ref{SD}),
the expression for the effective action $\Gamma$ simplifies, i.e.,
\begin{equation}
\Gamma= -i\,\mbox{Tr}\left[\mbox{Ln} G^{-1} +\frac{1}{2}\left(S^{-1}G-1\right)\right] +
\frac{i}{2}\left[\mbox{Ln} D^{-1} +D_{0}^{-1}D-1\right] .
\label{omega}
\end{equation}
Up to the one-loop photon contribution, which we will neglect in the following, this is formally the same result
as in Ref.~\cite{Gorbar2008PRB}. Therefore, we can use a similar derivation and
change only the form of the Green's function $G_{s}(u,u^\prime)$ in order to take into account
that a wave function renormalization is nontrivial here and that all dynamical parameters are
functions of the Landau index. We finally obtain
\begin{eqnarray}
\Omega &=&-\frac{1}{4\pi \ell^2} \sum_{\sigma=\pm1}
\Bigg\{
\sign(\mu_{0,\sigma})
\left(\mu_{0,\sigma}+\mu-\sigma \tilde{\Delta}_{0,\sigma}\right) \theta(|\mu_{0,\sigma}|
-|\tilde{\Delta}_{0,\sigma}|)
+\sign(\tilde{\Delta}_{0,\sigma})
\left[\tilde{\Delta}_{0,\sigma}-\sigma \left(\mu+ \mu_{0,\sigma}\right)\right]
\theta(|\tilde{\Delta}_{0,\sigma}|-|\mu_{0,\sigma}|)\nonumber\\
&+&2\sum_{n=1}^{\infty}  \left[
\sign(\mu_{n,\sigma}) \left(\mu_{n,\sigma}+\mu\right) \theta(|\mu_{n,\sigma}|-|E_{n,\sigma}|)
+\frac{E_{n,\sigma}^2+ 2 n f_{n,\sigma} \varepsilon_\ell^2}{E_{n,\sigma}}
\theta(|E_{n,\sigma}|-|\mu_{n,\sigma}|)-2 \sqrt{2n} \varepsilon_\ell\right]\Bigg\},
\label{Omega-2}
\end{eqnarray}
where we introduced the function $\theta(n-1)$, which is
defined so that $\theta(n-1)=0$ for $n= 0$ and $\theta(n-1)=1$ for $n\geq 1$.
At a non-zero temperature, we make the replacement $\omega\to i\omega_{m}\equiv i \pi T (2m+1)$
and use Matsubara sums instead of the frequency integrations, and obtain
\begin{eqnarray}
\Omega &=&-\frac{1}{4\pi \ell^2} \sum_{\sigma=\pm1}
\Bigg\{
\left(\mu_{0,\sigma}+\mu-\sigma \tilde{\Delta}_{0,\sigma}\right) \tanh\left(\frac{\mu_{0,\sigma}
-\sigma \tilde{\Delta}_{0,\sigma}}{2T}\right)\nonumber\\
&+&2\sum_{n=1}^{\infty}  \left[
\left(\mu_{n,\sigma}+\mu\right) \left[ n_F(E_{n,\sigma}-\mu_{n,\sigma})-n_F(E_{n,\sigma}
+\mu_{n,\sigma}) \right]-2\sqrt{2n}  \varepsilon_\ell\left[1-2n_F(\sqrt{2n}\varepsilon_\ell )\right]
\right.\nonumber\\
&&\left.\hspace{0.5in}+\frac{E_{n,\sigma}^2+ 2 n f_{n,\sigma} \varepsilon_\ell^2}{E_{n,\sigma}}
 \left[ 1-n_F(E_{n,\sigma}-\mu_{n,\sigma})-n_F(E_{n,\sigma}+\mu_{n,\sigma}) \right]\right]\Bigg\}.
\label{Omega-3}
\end{eqnarray}

\end{document}